\documentclass[
aps,
pra,
superscriptaddress,
showpacs,
showkeys,
notitlepage,
twocolumn, 
numbers,
longbibliography,
nofootinbib]
{revtex4-1}

\usepackage[T1]{fontenc} 
\usepackage[utf8]{inputenc} 
\usepackage{bbold}
\usepackage[english]{babel} 
\usepackage{graphicx, xcolor}
\usepackage{amsfonts, amssymb, amsmath, mathrsfs, calc, array, mathdesign, mathtools}
\usepackage[separate-uncertainty=true]{siunitx}
\usepackage{hyperref}

\def\bra#1{\mathinner{\langle{#1}|}}
\def\ket#1{\mathinner{|{#1}\rangle}}

\def\ontop#1#2{\setbox0\hbox{#2}\copy0\llap{\raise\ht0\hbox{#1}}}
\definecolor{darkblue}{rgb}{0,0,0.93} 
\definecolor{darkred}{rgb}{0.8,0,0} 
\usepackage{graphicx,trimclip,amsmath,wasysym}

\hypersetup{
colorlinks=true, 
linkcolor=darkblue, 
citecolor=darkred, 
filecolor=darkblue, 
urlcolor=darkblue
}

\def\NOTEPAB#1{{\textcolor{blue}{\bf [#1]}}}        

\begin{document}

\title{Quantum simulation of quantum relativistic diffusion via quantum walks}

\author{Pablo Arnault}
\email{pablo.arnault@ific.uv.es}
\affiliation{Departamento de F{\'i}sica Te{\'o}rica and IFIC, Universidad de Valencia and CSIC, Dr.\ Moliner 50, 46100 Burjassot, Spain}

\author{Adrian Macquet}
\affiliation{Univ.\ C{\^o}te d'Azur and Obs.\ C{\^o}te d'Azur, Lab.\ ARTEMIS, 35150 Janz{\'e}, France}

\author{Andreu Anglés-Castillo}
\affiliation{Departamento de F{\'i}sica Te{\'o}rica and IFIC, Universidad de Valencia and CSIC, Dr.\ Moliner 50, 46100 Burjassot, Spain}

\author{Iv{\'a}n M{\'a}rquez-Mart{\'i}n}
\affiliation{Departamento de F{\'i}sica Te{\'o}rica and IFIC, Universidad de Valencia and CSIC, Dr.\ Moliner 50, 46100 Burjassot, Spain}
\affiliation{Aix-Marseille Univ., Universit{\'e} de Toulon, CNRS, LIS, Marseille, France}

\author{Vicente Pina-Canelles}
\affiliation{Departamento de F{\'i}sica Te{\'o}rica and IFIC, Universidad de Valencia and CSIC, Dr.\ Moliner 50, 46100 Burjassot, Spain}

\author{Armando P{\'e}rez}
\affiliation{Departamento de F{\'i}sica Te{\'o}rica and IFIC, Universidad de Valencia and CSIC, Dr.\ Moliner 50, 46100 Burjassot, Spain}

\author{Giuseppe Di Molfetta}
\affiliation{Universit{\'e} publique, CNRS, LIS, Marseille, France}
\affiliation{Quantum Computing Center, Keio University}
\affiliation{Departamento de F{\'i}sica Te{\'o}rica and IFIC, Universidad de Valencia and CSIC, Dr.\ Moliner 50, 46100 Burjassot, Spain}

\author{Pablo Arrighi}
\affiliation{Aix-Marseille Univ., Universit{\'e} de Toulon, CNRS, LIS, Marseille and IXXI, Lyon, France}

\author{Fabrice Debbasch}
\affiliation{Sorbonne Universit{\'e}, Observatoire de Paris, Universit{\'e} PSL, CNRS, LERMA, F-75005, Paris, France}

\begin{abstract}
Two models are first presented, of one-dimensional discrete-time quantum walk (DTQW) with temporal noise on the internal degree of freedom (i.e., the coin): (i) a model with both a coin-flip and a phase-flip channel, and (ii) a model with random coin unitaries.
It is then shown that both these models admit a common limit in the spacetime continuum, namely, a Lindblad equation with Dirac-fermion Hamiltonian part and, as Lindblad jumps, a chirality flip and a chirality-dependent phase flip, which are two of the three standard error channels for a two-level quantum system.
This, as one may call it, Dirac Lindblad equation, provides a model of quantum relativistic spatial diffusion, which is evidenced both analytically and numerically.
This model of spatial diffusion has the intriguing specificity of making sense only with original unitary models which are relativistic in the sense that they have chirality, on which the noise is introduced: The diffusion arises via the by-construction (quantum) coupling of chirality to the position.
For a particle with vanishing mass, the model of quantum relativistic diffusion introduced in the present work, reduces to the well-known telegraph equation, which yields propagation at short times, diffusion at long times, and exhibits no quantumness.
Finally, the results are extended to temporal noises which depend smoothly on position.

\end{abstract}



\maketitle


\section{Introduction}

In classical continuous media theory, diffusion in the absence of force field
designates irreversible evolutions which are induced by and compensate for inhomogeneous repartitions of certain
extensive quantities (charge, particle number, momentum and energy density). Under diffusions,
the medium relaxes towards an equilibrium state where these quantities have time- and space-independent concentrations.
On the microscopic scale, diffusion is always associated with random motions. The simplest example is Brownian motion, 
which was first observed by Brown in 1827, revisited theoretically by Einstein in 1905 \cite{Einstein1905} and is now the cornerstone of modern 
stochastic process theory.
Diffusion also occurs in quantum and/or relativistic systems. The first attempts to describe quantum non-relativistic diffusion processes were made in the 1970's (see Ref.\ \cite{DC07b}). Quantum diffusion \cite{book_Percival, GZ00a, book_Schlosshauer} occurs in open quantum systems interacting with their environment and is usually described through a deterministic differential transport equation of the Lindblad form \cite{BP07} obeyed by the so-called reduced density operator of the system. 
The problem of finding macroscopic models of relativistic non-quantum diffusion was first considered in the 1940's by Landau and Eckhart \cite{LL87, Eckart1940}. It was first revisited by Cattaneo \cite{Cattaneo1949}, who 
suggested to model relativistic diffusion through the telegraph equation, and whose work, coupled with the Grad expansion technique \cite{Grad_1949} (see also Ref.\ [52] in Ref.\ \cite{I87a}), laid the basis of the so-called Extended Thermodynamics theories \cite{MR93a}. All models produced by these efforts present serious difficulties, which range from non-causality and instability (see Refs.  [6,7] and [49] in Ref.\ \cite{I87a}) to experimental refutation\footnote{This experimental refutation has been obtained, in the references cited, in the non-relativistic regime, the relativistic regime being more difficult to access. That being said, because this non-relativistic regime can be derived as a limit of relativistic Extended-Thermodynamics models, the abovementioned experiments refute at least the kind of limit which has been taken and at most these relativistic Extended-Thermodynamics models themselves.} (see Refs. [58--59] in Ref.\ \cite{I87a}). Also, all implementations of the Extended Thermodynamics philosophy are based on truncating the Grad expansion, which usually diverges. It therefore comes as no surprise that some experimental predictions of Extended Thermodynamics seem to diverge with the supposed precision of the implementation, thus making Extended Thermodynamics void of any real predictive power, at least for some phenomena like second sound (see the last chapter of \cite{MR93a}). 
The problem of finding microscopic models of relativistic non quantum diffusion was in theory entirely solved by writing down a relativistic version of Boltzmann equation \cite{DvL03a, DvL03b}, but practical computations and conceptual issues
necessitated also extending stochastic process theory to the relativistic realm. 
The first relativistic stochastic process was considered by Dudley \cite{D65a}. Though well-defined mathematically, this process is not of obvious 
physical usefulness and fails to predict important phenomena like thermalization. The first relativistic process of physical relevance is the Relativistic Ornstein Uhlenbeck Process (ROUP) and was presented in 1997 by Debbasch, Mallick and Rivet \cite{ROUP0, ROUPHydroLim, ROUPCurved}. Franchi and Le Jan then revisited the Dudley process taking into account the physics of the ROUP, both in flat and curved spacetimes \cite{FLJ04, FlJ07a}. A process mixing aspects of the Dudley process and of the ROUP was later introduced by Dunkel and H{\"a}nngi \cite{DH05a, DH05b}.
Finally, the ROUP served as a basis for the construction of the first macroscopic model of bounded velocity diffusion free of any physical and mathematical pathology \cite{DC07b, CD07d, ROUPUnifying1, ROUPUnifying2}. Models
of this type can be used for relativistic and non-relativistic bounded velocity diffusions \cite{DEF12a}.
Let us eventually mention two historical references on this topic of relativistic stochastic processes, Refs. \cite{Hakim1965, Hakim1968}.

In this paper, we develop a novel quantum-simulation scheme which models relativistic diffusive transport in the quantum regime, by mimicking an appropriate Lindblad equation via the continuum limit of a noisy discrete-time quantum walk (DTQW).
Quantum simulation is a flourishing field, thanks to its advantages with respect to classical simulation: classical computers are especially inefficient at simulating quantum dynamics of highly entangled systems.

The advantage of some quantum algorithms with respect to their classical counterparts is already known, as with Grover's algorithm \cite{Grover96}, which can solve the task of searching an element in a database quadratically faster than known classical algorithms. 
Grover's algorithm can be written in terms of a DTQW, whose spatial probability distribution spreads quadratically faster than that of a random walk. 
One more example is the proposal of using quantum walks for ranking nodes on a network \cite{Chawla2019}.
Another application of DTQWs is the direct simulation of physical dynamics: if the set-up for them is chosen appropriately, they can be used to model several physical phenomena, e.g., the dynamics of fermions in the free case \cite{BB94a} or in an external Abelian (i.e., electromagnetic) \cite{BB94a, AD15, AD16} or non-Abelian \cite{ADMDB16} Yang-Mills gauge field, neutrino flavor oscillations \cite{Molfetta2016}, and fermion confinement \cite{MrquezMartn2017}.
These DTQW schemes are not limited to square-lattices backgrounds, but can also be designed on triangle and honeycomb lattices \cite{ADMMMP18a, Jay2019}.
Moreover, the (classical) field dynamics that DTQWs can mimick is not limited to flat-spacetime backgrounds, can be extended to curved spacetimes \cite{DMD14,AD17,AF17,ADMF16,Arrighi2019}.
Action principles for DTQWs have been suggested, and the spacetime covariance of the latter has been investigated, both in flat \cite{Arrighi2014, Debbasch2019} and in curved \cite{Debbasch2019b} spacetime.

The connections between DTQWs and lattice gauge theories have also been explored \cite{MMADMP2018, APAF18, Cedzich2019, DMA19}, and Wigner functions for DTQWs
have been proposed in Refs. \cite{Hinarejos_2013, Hinarejos_2015, DebbaschWigner19}.
A crucial feature of DTQWs is that they are intrinsically causal, i.e., information propagates, at most, at a finite velocity $c=1$, which is why DTQWs are \emph{a priori} especially suited to model
quantum relativistic diffusions. 

This article is organized as follows. Basics about DTQWs are reviewed in Sec.\ \ref{sec:basics}, while Sec.\ \ref{sec:noise} introduces two models with temporal noise and
a common continuous limit of the Lindblad form. Section \ref{sec:QRD} explores the phenomenology of this limit. Section \ref{sec:spatiotemp} extends the previous results to temporal noises which depend smoothly on space. All results are summarized and discussed in the final section, while technicalities are dealt with in the Appendices.

%

\section{The unitary model: discrete-time quantum walk on the line}
\label{sec:basics}

\subsection{Presentation}

Consider a series of quantum states, $\ket{\Psi_t}$, indexed by the discrete time $t\in \mathbb{N}\epsilon$, where $\epsilon > 0$ is the time step, and belonging to a Hilbert space $\mathcal{H}_{\text{c}} \otimes \mathcal{H}_{\text{p}}$, where (i) $\mathcal{H}_{\text{c}}$ is the so-called \emph{coin (Hilbert) space}, which is two dimensional and accounts for an internal, two-state degree of freedom (hereafter d.o.f.) we call coin (hence the index ``c''), and (ii) $\mathcal{H}_{\text{p}}$ is the \emph{position (Hilbert) space} (hence the index ``p'').
The probability amplitudes of this state on the position basis, $\{ \ket{x}, x \in \mathbb{Z}a\}$, where $a>0$ is the lattice spacing, are thus described by a two-component wave function, $\Psi_{t,x} \equiv  \langle x |  \Psi_t \rangle \equiv ( \psi^{L}_{t,x}, \psi^{R}_{t,x} )^{\top} $, where $\top$ denotes the transposition.

Consider that $\ket{\Psi_t}$ evolves according to the following standard model of discrete-time quantum walk on the line,
\begin{equation} \label{eq:evol}
\ket{\Psi_{t+\epsilon}} = \hat{U}_{\hat{\xi}^0_t,\hat{\xi}^1_t,\hat{\theta}_t,\hat{\chi}_t} \ket{\Psi_{t}} \, ,
\end{equation}
where the one-step evolution operator, called \emph{walk operator}, is
\begin{equation} \label{eq:evolution_op}
\hat{U}_{\hat{\xi}^0_t,\hat{\xi}^1_t,\hat{\theta}_t,\hat{\chi}_t} \equiv C_{\hat{\xi}^0_t,\hat{\xi}^1_t,\hat{\theta}_t,\hat{\chi}_t} S(\hat{p}) \, .
\end{equation}
This evolution operator is the succession of two unitary operators. 

The first one is a coin-dependent \emph{shift operator},
\begin{equation} \label{eq:shift}
S(\hat{p}) \equiv \begin{bmatrix}
e^{\mathrm{i} a\hat{p}} & 0 \\
0 &  e^{-\mathrm{i} a\hat{p}}
\end{bmatrix} = e^{\mathrm{i}a\hat{p} \sigma^3} \, ,
\end{equation}
where $\sigma^i$ is the $i$th  Pauli matrix, and $\hat{p}$ is the quasimomentum operator, which is Hermitian (this ensures that $S(\hat{p})$ is unitary) and satisfies $e^{\mathrm{i}a\hat{p}} = \sum_{x} \ket{x} \! \! \bra{x+a}$, so that the upper (resp.\ lower) component, $\psi_t^{L}$ ($\psi_t^{R}$), is shifted left (resp.\ right), hence the superscript $L$ (resp.\ $R$). 
Notice that we have implicitly introduced the $LR$ basis of the coin space, namely, $(\ket{L}, \ket{R})$, which we have identified with $((1,0)^{\top}, (0,1)^{\top})$.

The second operator is a so-called \emph{coin operator},
\begin{align} \label{eq:coin_op}
C_{\hat{\xi}^0_t,\hat{\xi}^1_t,\hat{\theta}_t,\hat{\chi}_t} &\equiv 
e^{\mathrm{i} \hat{\xi}^0_t} \begin{bmatrix}
e^{\mathrm{i} \hat{\xi}^1_t} \cos \hat{\theta}_t & \mathrm{i} e^{\mathrm{i} \hat{\chi}_t} \sin \hat{\theta}_t \\
\mathrm{i} e^{-\mathrm{i} \hat{\chi}_t} \sin \hat{\theta}_t & e^{-\mathrm{i} \hat{\xi}^1_t} \cos \hat{\theta}_t
\end{bmatrix}  \, ,
\end{align}
which is nothing but an arbitrary $2 \times 2$ unitary matrix with $4$ operator-valued entries,
\begin{equation}
\hat{f}_t \equiv (\hat{f}^l_t)_{l=0,.\hspace{-0.5pt}.\hspace{-0.5pt}.,3} \equiv (\hat{\xi}^0_t, \hat{\xi}^1_t,\hat{\theta}_t,\hat{\chi}_t) \, ,
\end{equation}
acting on the position space.
%
 %
To endow the $\hat{f}^l_t$s (varying $l$ and $t$), with the highest degree of arbitrariness, one must consider them (i) Hermitian, to ensure the \emph{unitarity} of the coin operator, and (ii) \emph{diagonal in the position basis}, that is, 
\begin{equation}
\hat{f}^l_t\ket{x} \equiv f^l_{t,x}\ket{x} \, ,
\end{equation}
(which defines the sequences $f^l: (t,x) \mapsto f^l_{t,x}$, which are real-valued because of Hermiticity), so as to ensure the \emph{locality} of the walk operator in position space.

The $\hat{f}^l_t$s being diagonal in the position basis, they commute between each other, so that we can write the coin operator as
\begin{align}
C_{\hat{\xi}^0_t,\hat{\xi}^1_t,\hat{\theta}_t,\hat{\chi}_t} = 
e^{\mathrm{i} \hat{\xi}^0_t}  e^{\mathrm{i}\frac{\hat{\chi}_t}{2} \sigma^3} e^{\mathrm{i}\frac{\hat{\xi}^1_t}{2} \sigma^3} e^{\mathrm{i}\hat{\theta}_t \sigma^1}  e^{\mathrm{i}\frac{\hat{\xi}^1_t}{2}\sigma^3}e^{-\mathrm{i}\frac{\hat{\chi}_t}{2}\sigma^3} \, . \label{eq:chi}
\end{align}
This readily shows that, if $\chi$ is space independent, it simply codes for a global change of coin basis at time $t$, which, in addition, commutes with the coin-dependent shift operator, so that, if $\chi$ is moreover time independent, it does not affect the dynamics.
In Appendix \ref{app:Euler_angles}, we explain the reasons for choosing this parametrization, Eq.\ (\ref{eq:coin_op}) (or (\ref{eq:chi})), for the unitary group.
%

When the entries of the coin operator are time and space independent, the behavior of this dynamical system, Eq.\ (\ref{eq:evol}), is well know.
It yields, whatever the values of the entries, two propagation fronts, one to the left, and the other to the right, and thus exhibits, in particular, ballistic spread, i.e., $O(t)$ spread. 
In the long-time limit, the spread is exactly $\sigma_{\infty}(t) = (a/\epsilon)t \sqrt{1-\sin \theta}$  \cite{Chandrashekar2010,Prez2016}.
Notice in particular that this spread\footnote{In Ref.\ \cite{Prez2016}, the spread is computed for $\xi^0=\xi^1=0$ and $\chi=\pi/2$, but one can adapt the demonstration to arbitrary values for these angles. In short: (i) a constant $\chi$ does not even intervene in the dispersion relation of the DTQW, (ii) a constant $\xi^0$ does not intervene in the group velocity of the DTQW, and (iii) a constant $\xi^1$ is just a constant shift in the Brillouin zone, which is irrelevant when performing integrals of functions which are periodic with period the size of the Brillouin zone. For a rigorous mathematical proof of the long-time probability distribution of the DTQW (from which one can of course compute, in particular, the variance), see \cite{Konno2002}.} is independent of $\xi^0$, $\xi^1$ and $\chi$.

Notice that we use hats for operators acting on the position space, the reason for this being that we do not identify them with their matrix representation. In contrast, we do not use hats for operators acting on the coin space, the reason for this being that we do identify them with their matrix representation.

\subsection{Continuum limit} \label{subsec:Continuum_limit}

It is well known \cite{DMD14} (i) that the above lattice model, Eq.\ (\ref{eq:evol}), possesses a continuum limit, $\epsilon \rightarrow 0$ and $a \rightarrow 0$, for the ballistic scaling\footnote{Ballistic scaling means $\epsilon \propto a$, and we can choose $\epsilon=a$ without loss of generality, i.e., setting $a/\epsilon$ as the speed unit.},
\begin{equation}
a = \epsilon \, ,
\end{equation}
(which we assume from now on when taking continuum limits), provided, essentially, that $\xi^0$, $\xi^1$ and $\theta$, also go to zero with $\epsilon$, and (ii) that the richest situation \cite{DMD14} is obtained when they scale as $\epsilon$, i.e.,
\begin{subequations}
\begin{align}
\xi^0 &\equiv \epsilon \bar{\xi}^{0} \\
\xi^1 &\equiv \epsilon \bar{\xi}^{1} \\
\theta &\equiv \epsilon \bar{\theta} \, ,
\end{align}
\end{subequations}
which we assume from now on when taking continuum limits (it will be recalled), where $\bar{\xi}^0$, $\bar{\xi}^1$ and $\bar{\theta}$, are arbitrary functions of time and space.
Indeed, when these conditions are satisfied\footnote{Together with regularity conditions for the quantities that we Taylor expand in $\epsilon$ \cite{DMD14}.}, the evolution operator reads
\begin{equation}
\hat{U}_{\epsilon \hat{\bar{\xi}}^0,\epsilon \hat{\bar{\xi}}^1,\epsilon \hat{\bar{\theta}},\hat{\chi}} =  \mathbb{1} - \mathrm{i} \epsilon H_{\hat{\bar{\xi}}^0,\hat{\bar{\xi}}^1,\hat{\bar{\theta}}, {\hat{\chi}}}(\hat{p}) + O(\epsilon^2) \, ,
\end{equation}
so that it has a valid continuum limit --  i.e., the walk operator tends to the identity when $\epsilon$ tends to zero -- which is generated by the following Hamiltonian,
\begin{equation}
H_{\hat{\bar{\xi}}^0,\hat{\bar{\xi}}^1,\hat{\bar{\theta}},\hat{\chi}}(\hat{p}) \equiv \alpha^1 (\hat{p} + \hat{\bar{\xi}}^1) + M_{\hat{\bar{\theta}},\hat{\chi}} \alpha^0 - \hat{\bar{\xi}}^0 \mathbf{1}_2 \, ,
\end{equation}
which is a generalization of the 1D Dirac Hamiltonian for a particle with mass matrix
\begin{equation}
M_{\hat{\bar{\theta}},\hat{\chi}} \equiv -\hat{\bar{\theta}} D(\hat{\chi}) \, ,
\end{equation}
where
\begin{equation}
D(\hat{\chi}) \equiv \text{diag}(e^{\mathrm{i}\hat{\chi}},e^{-\mathrm{i}\hat{\chi}}) \, ,
\end{equation}
and charge $q=-1$, coupled to an electromagnetic potential with covariant components 
\begin{subequations}
\begin{align}
A_0 &\equiv \bar{\xi}^0 \\
A_1 &\equiv -\bar{\xi}^1 \, ,
\end{align}
\end{subequations}
and with the following representation of the alpha matrices,
\begin{subequations}
\begin{align}
\alpha^0 &\equiv  \sigma^1\\
\alpha^1 &\equiv -\sigma^3 \, .
\end{align}
\end{subequations}
We have also introduced $\mathbf{1}_2$, the $2\times 2$ identity matrix.

Assume $\chi \neq 0$: even if $\chi$ is spacetime independent, $D(\hat{\chi})$ cannot be absorbed in $\alpha^1$ because $D(\hat{\chi})^2 \neq \mathbf{1}_2$, so that the Clifford algebra is not be satisfied, and so the resulting Dirac equation does not square to the Klein-Gordon equation.
When $\chi=0$, we recover, of course, a standard Dirac Hamiltonian with real (though possibly spacetime-dependent) mass,
\begin{equation}
m\equiv-\bar{\theta} \, ,
\end{equation}
namely,
\begin{subequations} \label{eq:HDirac}
\begin{align}
H_{\hat{A}_0,-\hat{A}_1,\hat{m},\hat{\chi}=0}(\hat{p}) &= H^{\text{Dirac}}_{\hat{m},\hat{A}_0}(\hat{p} -\hat{A}_1) \\
&\equiv \alpha^1 (\hat{p} -  \hat{A}_1) + \hat{m} \alpha^0 - \hat{A}_0 \mathbf{1}_2 \, . \label{eq:real_mass_Dirac}
\end{align}
\end{subequations}
For this reason, we assume from now on that $\chi=0$, and introduce, for the purpose of compactness of notations in the continuum-limits sections to come, $\bar{\chi} \equiv \chi/\epsilon$, so that,
\begin{equation}
\chi \equiv \epsilon \bar{\chi} = 0 \, .
\end{equation}
Notice that we choose $\chi=0$ solely for the sake of simplicity, -- i.e., to match with the standard Dirac Hamiltonian with \emph{real} (though possibly spacetime-dependent) mass, Eq.\ (\ref{eq:real_mass_Dirac}) --, that is, one could perfectly consider an arbitrary spacetime dependence for $\chi$ in the computations to come, without any change in the results but that one\footnote{Regarding the role played by a spacetime-\emph{independent} $\chi$ for the above Dirac Hamiltonian, Eq.\ (\ref{eq:HDirac}), the reader may be interested in Ref.\ \cite{Trzetrzelewski2015}.
Regarding the role played by a spacetime-\emph{dependent} $\chi$ in another class of continuum limits, the reader may be interested in Ref.\ \cite{DMD14}.
Regarding the role played by the four spacetime-dependent angles in the original, spacetime-lattice model, Eq.\ (\ref{eq:evol}), the reader may consult, for $\xi^0$ and $\xi^1$, Refs. \cite{DMD14,AD16,MMADMP2018, APAF18, Cedzich2019, D19a}, -- which show, among other results and in various, related settings, that these two angles correspond to lattice versions of the electromagnetic potential, having lattice U(1) gauge invariance --, and, for $\theta$ and $\chi$, Ref.\ \cite{D19a}, which shows that these two angles encode the curvature of a discrete spacetime.}.

\section{Two models of temporal coin noise with a common continuum limit} \label{sec:noise}

\subsection{Discrete-time quantum walk with coin-flip and phase-flip channels} \label{subsec:coin-depol}

\subsubsection{Lattice model}

A simple and well-known model of temporal coin noise for the DTQW introduced in Eq.\ (\ref{eq:evol}), is to consider that, for each evolution $t \rightarrow t+\epsilon$, the walker follows the unitary evolution with some probability $1-\pi_+$, with $\pi_+$ independent on time, and that, with probability  $\pi_+=\pi_1+\pi_2$,
it undergoes either a phase-flip channel, that is, a coin-dependent phase flip\footnote{This phase flip is coin dependent, but the specification ``coin dependent'' (i.e., in general, ``internal-state dependent'') can be omitted, since a coin-independent phase flip leads to a trivial, identity channel, which is rarely of interest.}, i.e., evolves through the unitary $\sigma^3$, with probability $\pi_1$, or a bit-flip channel, that is, a bit flip, i.e., evolves through the unitary $\sigma^1$, with probability $\pi_2$ \cite{Preskill_course,Cappellaro_course}.

To describe the behavior of a quantum noisy system statistically, i.e., its average behavior over a large number of realizations of the noisy dynamics, one needs the density-operator formalism.
%
In the present case, the evolution equation for the density operator, $\hat{\rho}_t$, is simply,
{\small
\begin{equation} \label{eq:depol_channels}
\hat{\rho}_{t+\epsilon} = (1 - \pi_1 - \pi_2)  \hat{U}_{ \hat{{f}}_{t}} \hat{\rho}_t \hat{U}^{\dag}_{ \hat{{f}}_{t}} + {\pi_1} \sigma^3 \hat{\rho}_t  \sigma^3  + {\pi_2} \sigma^1 \hat{\rho}_t  \sigma^1 \, .
\end{equation}}

\subsubsection{Continuum limit}

A simple condition for Eq.\ (\ref{eq:depol_channels}) taken for $\hat{f}_t=\epsilon \hat{\bar{f}}_t$ to have a formal continuum limit as $\epsilon \rightarrow 0$ is to assume that $\pi_l \rightarrow_{\epsilon\rightarrow 0} 0$, $l=1,2$. For simplicity, we assume that they scale as $\epsilon$,
\begin{equation}
\pi_l \equiv \epsilon \tilde{\pi}_l  \, , \ \ l=1,2 \, ,
\end{equation}
where $\tilde{\pi}_l$ is an arbitrary real number corresponding to a probability per unit time. After Taylor expanding Eq.\ (\ref{eq:depol_channels}) at first order in $\epsilon$, cancelling out the zeroth-order terms, and letting $\epsilon \rightarrow 0$, we are lead to the following equation,
\begin{equation} \label{eq:Lindblad_eq}
\partial_{t} \hat{\rho} = - \mathrm{i}  [\hat{H}^o,\hat{\rho}]+ \mathcal{L}_{\tilde{\Pi}}(\hat{\rho})  \, ,
\end{equation}
where the Hamiltonian part is the Dirac one, see Sec.\ \ref{subsec:Continuum_limit},
\begin{equation} \label{eq:Ho}
\hat{H}^o_t \equiv H^{\text{Dirac}}_{\hat{m},\hat{A}_0|_t}(\hat{p} -\mathrm{i}\hat{A}_1|_t) \, ,
\end{equation}
and the non-Hamiltonian, but still trace-preserving one, reads,
\begin{align} \label{eq:Lind}
\mathcal{L}_{\tilde{\Pi}}(\hat{\rho}_t) 
\equiv \tilde{\pi}_1 \left[ \sigma^3 \hat{\rho}_t \sigma^3 - \hat{\rho}_t \right]  + \tilde{\pi}_2 \left[ \sigma^1 \hat{\rho}_t \sigma^1 - \hat{\rho}_t \right] \, ,
\end{align}
and can be recast in a Lindblad form, whose most general writing is
\begin{equation} \label{eq:Lindblad_op}
\mathcal{L}_{X}(\hat{\rho}_t) \equiv \sum_{i\in I} X_i \left[ {L}_i \hat{\rho}_t {L}_{i}^{\dag} - \frac{1}{2} \{ L_i^{\dag} L_i, \hat{\rho}_t \} \right] \, ,
\end{equation}
where  $X \equiv (X_i)_{i\in I}$ is an arbitrary family of non-negative real numbers indexed by the label $i$ belonging to some indexing space $I$, and  the $L_i$s are the so-called Lindblad or jump operators, which act on the Hilbert space of the system and can be non-Hermitian.
In the present case, i.e., in Eq.\ 
(\ref{eq:Lind}), we have $X=\tilde{\Pi}\equiv(\tilde{\pi}_1,\tilde{\pi}_2)$, $i=l=1,2$, and two Lindblad operators,
\begin{subequations} \label{eq:Lls}
\begin{align}
L_1 &\equiv \sigma^3 \\
L_2 &\equiv \sigma^1 \, ,
\end{align}
\end{subequations}
which are Hermitian, act solely on the coin space, and whose square is proportional to $\mathbf{1}_2$.

\subsection{Discrete-time quantum walk with random coin unitaries}
\label{subsec:random-unitaries}

\subsubsection{Lattice model}

Another simple and well-known model of temporal coin noise for the DTQW introduced in Eq.\ (\ref{eq:evol}), is to consider that, for each evolution $t \rightarrow t+\epsilon$, the values of the coin-operator parameters are not fixed numbers but sampled from respective probability distributions \cite{DiMolfetta2016}, so that we denote them with a prime, ${\xi^0}'_{\! \! \! t,x}$,  ${\xi^1}'_{\! \! \!  t,x}$, $\theta'_{t,x}$, and $\chi'_{t,x}$. 
For simplicity, we assume that these random values can depend on space only through their mean value, i.e., they have space-independent fluctuations and thus respective centered probability distributions (that is why we speak of \emph{temporal} noise), that we denote $p^l_t$. 
This means that
\begin{align} \label{eq:omegaa}
{f'}^l_{\! \! \! t,x} &\equiv \epsilon \bar{f}^{l}_{t,x} + \omega^l_t \ \text{for} \ l=0,...,3 \, ,
\end{align}
where (i) for $l=0,...,2$ (resp. $l=3$), $\epsilon \bar{f}^{l}_{t,x}$ (resp. $0$) is the mean value, which we have assumed scaling as $\epsilon$ (resp, vanishing), in order to recover, in the noiseless case, the previous Hamiltonian evolution, Eq.\ (\ref{eq:HDirac}), and (ii) ${\omega}^l_t \in \mathbb{R}$ is the space-independent fluctuation, newly sampled from the probability distribution $p^{l}_t$ at each time, and associated to a random variable $\Omega^l_t$.
Giving oneself a function $p^l_t$ of $\omega_t \in \mathbb{R}$ for each $t$ means assuming that the noise has \emph{temporal independence}, i.e., the random variables $\Omega_t$ and $\Omega_{t'}$ are independent\footnote{This implies, in particular, that the noise is classically Markovian, since one can give, for any $t$ and any $\omega_t\in \mathbb{R}$, the probability that $\Omega_t = \omega_t$, without the need to know the past history, i.e., the values $(\omega_{t'})_{\{t' < t \}}$, and hence (sufficient condition for Markovianity), without the need to know $(\omega_{t'})_{\{t' < t-\epsilon \}}$.} for $t'\neq t$.
Now, in addition to temporal independence, we assume \emph{stationarity}, i.e., that the $p^l_t$s do not depend explicitly on time: $p^l_t = p^l$.
The four time-indexed random variables $\Omega^l_t$ associated to the possible values $\omega^l_t$, are considered statistically independent, so that the probability density of getting $\omega^0_t$ and $\omega^1_t$ and $\omega^2_t$ and $\omega^3_t$ is given by the product $\prod_{l=0}^3 p^{l}(\omega^l_t)$.

The above model translates into the following evolution for the density operator, $\hat{\rho}_t$,
\begin{equation} \label{eq:evol_rho}
\hat{\rho}_{t+\epsilon} = \int  \! \! d\mu \ \mathcal{U}_{\epsilon \hat{\bar{f}}_{t} + {\omega}_t}\!(\hat{\rho}_t) \, ,
\end{equation}
where the integration measure is
\begin{equation} \label{eq:measure}
d\mu \equiv \prod_{l=0}^3 d\omega^l_t 	\, p^{l}(\omega^l_t) \, ,
\end{equation}
with the normalization condition,
\begin{equation}
\int \! \! d\mu = 1 \, ,
\end{equation}
and where each random unitary is given by
\begin{equation} \label{eq:one-step}
\mathcal{U}_{\epsilon \hat{\bar{f}}_{t} + {\omega}_t}\! (\hat{\rho}_t) \equiv \hat{U}_{\epsilon \hat{\bar{f}}_{t} + {\omega}_t} \hat{\rho}_t  \hat{U}^{\dag}_{\epsilon \hat{\bar{f}}_{t} + {\omega}_t} \, .
\end{equation}
We have omitted, for the sake of simplicity, the multiplication of $\omega_t$ by the identity operator acting on the position space, and will do so from now on unless otherwise mentioned.
As expected (since we impose the linearity of the theory and the Hermiticity of $\hat{\rho}_t$ \cite{book_Barnett}), this evolution, Eq.\ (\ref{eq:evol_rho}), has the form of a Kraus decomposition, the densities of the Kraus operators being simply the random unitaries $\hat{U}_{\epsilon \hat{\bar{f}}_{t} + {\omega}_t}$ (for simplicity, we have left the probability density in the integration measure $d\mu$, i.e., we have not included it in the definition of the Kraus operators).

\subsubsection{Continuum limit}
\label{subsubsection:cont-lim}

As in previous sections, we assume that the random variables introduced above, $\Omega_t^l$, actually result from the product $\Omega_t^l \equiv \phi^l(\epsilon) \tilde{\Omega}_t^l$, where $\phi^l(\epsilon)$ is a function going to zero with $\epsilon$, and \emph{$\tilde{\Omega}^l_t$ is new random variable that we introduce}.
This assumption ensures that $\hat{U}_{\epsilon \hat{\bar{f}}_{t} + {\Omega}_t}{\rightarrow}  \mathbb{1}$ as $\epsilon \rightarrow 0$, which, in turn, ensures that Eq.\ (\ref{eq:evol_rho}) remains consistent in that limit.
For a large class of functions $\phi^l$, we have that $\phi^l(\epsilon)$ is dominated, for $\epsilon \rightarrow 0$, by a term which scales as $\epsilon$ to some power $\nu_l>0$. 
One can show that only $ \nu_l = 1/2$ for all $l$s, delivers a non-trivial, non-unitary, trace-preserving limit for Eq.\ (\ref{eq:evol_rho}).
We thus assume, in the end,
\begin{equation} \label{eq:omega_scaling}
\Omega^l_t \equiv  \sqrt{\epsilon} \, \tilde{\Omega}^l_t \, , \ \ l=0,\hspace{-0.5pt}.\hspace{-0.5pt}.\hspace{-0.5pt}.,3 \, ,
\end{equation}
with $\tilde{\Omega}_t^l$ independent from $\epsilon$, i.e., with $\tilde{p}^l$ independent from $\epsilon$ \footnote{Notice that $\tilde{p}^l$ cannot depend on time since $p^l$ does not.}.
This assumption is crucial for the upcoming derivation to be valid, because for the Taylor expansion of Eq.\ (\ref{eq:Taylor1}) to hold.
In other words, this means that we have modified our model: indeed, for, e.g., one realization of this random-unitaries model, one samples, at each time, the $4$ values $\tilde{\omega}_t^l$ from their respective probability distributions $\tilde{p}^l$, and \emph{multiplies them} by $\sqrt{\epsilon}$ before taking the resulting products as arguments of the evolution operator\footnote{
That being said, maybe one can show that, though this condition (i.e., Eq.\ (\ref{eq:omega_scaling}) with $\tilde{\Omega}_t^l$ independent of $\epsilon$), is sufficient to derive the results to come, namely, Eq.\ (\ref{eq:Taylor2}), it is not necessary.
Indeed, maybe one can show that the following, \emph{milder} condition is sufficient to obtain Eq.\ (\ref{eq:Taylor2}), namely, that the random variables $\Omega_t^l$ scale as $\sqrt{\epsilon}$ only \emph{on average}, i.e., that their standard deviation does.
While the assumption made for the road we are going to follow, i.e., Eq.\ (\ref{eq:omega_scaling}) with $\tilde{\Omega}_t^l$ independent of $\epsilon$, implies the previous condition, the converse does not hold: indeed, although we can for sure always define $W_t^l \equiv \Omega_t^l/\sqrt{\epsilon}$, there is no reason, in the general case, that $W_t^l$ does not depend on $\epsilon$.
In the case of uniform or Gaussian distributions, the two conditions are equivalent, but their are a priori not in the general case.}.
This implies that the probability measure is not anymore that of Eq.\ (\ref{eq:measure}),  but that associated to the new random variable $\tilde{\Omega}^l_t $, namely,
\begin{equation} \label{eq:measuretilde}
d\tilde{\mu} \equiv \prod_{l=0}^3 d\tilde{\omega}^l_t 	\, \tilde{p}^{l}(\tilde{\omega}^l_t) \, .
\end{equation}

The Taylor expansion of $\hat{U}_{\epsilon \hat{\bar{f}}_{t} + \sqrt{\epsilon}{\tilde{\omega}}_t}$ at order $\epsilon$ is not completely trivial, but can be derived from Eq.\ (\ref{eq:chi}), and be cast as
\begin{align} \label{eq:Taylor1}
\hat{U}_{\epsilon \hat{\bar{f}}_{t} + \sqrt{\epsilon}{\tilde{\omega}}_t} =  1&-\mathrm{i}\epsilon \hat{H}^{o}_t  \\
&+ \mathrm{i} \sqrt{\epsilon} \sum_{l=0}^2  \tilde{\omega}^l_t {L}_{l} \nonumber   \\
&- \frac{1}{2} (\sqrt{\epsilon})^2 \sum_{l=0}^2  (\tilde{\omega}^l_t)^2 (L_{l})^2 \nonumber \\
&- (\sqrt{\epsilon})^2 \left[  \tilde{\omega}^0_t ( \tilde{\omega}^1_t \sigma^3 + \tilde{\omega}^2_t \sigma^1) +  \tilde{\omega}^2_t \tilde{\omega}^3_t (\mathrm{i} \sigma^2) \right] \nonumber \\
&+ O(\epsilon^{3/2}) \nonumber \, .
\end{align} 
In this Taylor expansion, we recover a known, Hamiltonian part, $\hat{H}^o_t$, given by Eq.\ (\ref{eq:Ho}), and the ${L}_{l}$s, $l=0,1,2$, are defined by Eqs.\ (\ref{eq:Lls}) with $L_0 \equiv  \mathbf{1}_2$.
Notice that the variable $\tilde{\omega}^3_t \equiv \tilde{\chi}_t$ only appears in the crossed terms, which are those just before the $O(\epsilon^{3/2})$.

Inserting the above Taylor expansion, Eq.\ (\ref{eq:Taylor1}), in the evolution equation, Eq.\ (\ref{eq:evol_rho}), and taking into account that the $\tilde{\Omega}_t^l$s, varying $l$, are \emph{independent} random variables (which is visible in the integration measure, Eq.\ (\ref{eq:measuretilde})), and all have \emph{vanishing mean}, yields
{\small
\begin{align}
&\hat{\rho} + \epsilon \partial_{t} \hat{\rho} = \hat{\rho}  + \epsilon \bigg[ - \mathrm{i}  [\hat{H}^o,\hat{\rho}] + \mathcal{L}_{\tilde{\Delta}^{\!2}}(\hat{\rho})  \bigg]  +  O(\epsilon^{3/2})  \, , \label{eq:Taylor2}
\end{align}}
the non-Hamiltonian term being given by Eq.\ (\ref{eq:Lind}), notice the abscence of $L_0$, with 
\begin{equation}
\tilde{\Delta}^{\!2}\equiv(\tilde{\delta}_l^2)_{l=0,.\hspace{-0.5pt}.\hspace{-0.5pt}.,3}
\end{equation}
instead of $\tilde{\Pi}\equiv(\tilde{\pi}_1,\tilde{\pi}_2)$, where $\tilde{\delta}_l$ is by definition the standard deviation of $\tilde{\Omega}_t^l$ for any $t$.
Because $\tilde{p}^l$ is time independent, all its moments are, and in particular $\tilde{\delta}_l^2$.
Cancelling out, in the previous equation, the zeroth-oder terms in $\epsilon$,
dividing then by $\epsilon$, and letting $\epsilon \rightarrow 0$, yields Eq.\ (\ref{eq:Lindblad_eq}) with $\tilde{\Delta}^{\!2}$ instead of $\tilde{\Pi}$.
\emph{Notice that neither the noise on $\xi^0$, nor that on $\chi$, have any effect in this continuum limit}.

Notice that the present random-unitaries-model formal continuum limit is for a DTQW that accepts a 1-step continuum limit \cite{DMD13b}.
The random-unitaries-model formal continuum limit has started to be explored for DTQWs accepting, not a 1-step continuum limit, but a 2-step one, in Ref.\ \cite{DiMolfetta2016}.

\section{Dirac Lindblad equation with chirality-flip channel: a model of quantum relativistic diffusion}
\label{sec:QRD}

\subsection{Description of the problem}

\subsubsection{Presentation} \label{subsubsec:presentation}

In the previous section, we have presented two spacetime-lattice models of quantum transport with temporal noise, Eqs.\ (\ref{eq:depol_channels}) and (\ref{eq:evol_rho}), that deliver, in the continuum limit, the same (1+1)D Lindblad equation with Dirac Hamiltonian part and two standard error channels on the chirality: (i) a phase-flip channel with rate (probability per unit time) $\gamma_1/2 = \tilde{\pi}_1 = \tilde{\delta}_1^2$, and (ii) a bit-flip channel with rate
$\gamma_2/2 = \tilde{\pi}_2 = \tilde{\delta}_2^2$.
This Lindblad equation reads
\begin{equation} \label{eq:Lindblad_eq_2}
\partial_{t} \hat{\rho} = - \mathrm{i}  [\hat{H}^o,\hat{\rho}]+ \mathcal{L}_{\Gamma/2}(\hat{\rho})  \, ,
\end{equation}
where  $\hat{H}^o$ is given by Eq.\ (\ref{eq:Ho}), $\mathcal{L}_{X}(\hat{\rho})$ by Eq.\ (\ref{eq:Lindblad_op}), and
\begin{equation}
\Gamma \equiv (\gamma_1,\gamma_2) \, .
\end{equation}
For the sake of simplicity, we choose a vanishing electric potential,
\begin{equation} \label{eq:vanishing em pot}
(A_0,A_1)=0 \, ,
\end{equation}
and a mass term
\begin{equation} \label{sent:mass_term}
m \ \text{independent of both space and time}.
\end{equation}

How does the noise,  $\mathcal{L}_{\Gamma/2}(\hat{\rho})$, on the chirality d.o.f. of the Dirac fermion, affect the dynamics of the spatial d.o.f.?
A coupling between these two d.o.f.s is indeed expected, at a quantum level, because the mass entangles them
\footnote{
This does not hold in a non-relativistic setting, i.e., the massive, free dynamics, does not entangle the internal and external d.o.f.s. The presence of a \emph{non-uniform} magnetic field does always produce, be the model relativistic or not, such an entanglement, which is exemplified by the historical Stern-Gerlach experiment \cite{Roston2005}, which can be accounted for by a non-relativistic model.}.
We will see in Sec.\ \ref{subsec:telegraph} that, although a vanishing mass indeed destroys the \emph{purely quantum} coupling (i.e., the entanglement) between the internal and external d.o.f.s, the relativistic nature of the equation still introduces a certain coupling between the internal and external d.o.f.s, but which can be seen as \emph{purely classical}, i.e., the intrinsic quantum nature of the chirality d.o.f. has no purely quantum phenomenal consequence, and this chirality could, in this massless case, be described in a non-quantum manner.

\subsubsection{Equations on the Pauli basis}

We decompose, for convenience, $\hat{\rho}_t$ on the Pauli basis,
\begin{equation} \label{eq:decomp}
\hat{\rho}_t = \frac{1}{2} \sum_{\mu=0}^3  \hat{r}_t^{\mu} \sigma^{\mu} \, ,
\end{equation}
where $\sigma^0\equiv\mathbf{1}_2$.
The $\hat{r}_{t}^{\mu}$s are observables acting solely on the position space, which can be obtained from $\hat{\rho}_t$ by the following partial trace, denoted $\text{Tr}_{\mathrm{c}}$, on the internal d.o.f.,
\begin{equation}  \label{eq:Tr}
\hat{r}_{t}^{\mu}  =  \text{Tr}_{\mathrm{c}} (\hat{\rho}_t \sigma^{\mu} ) \, .
\end{equation}
In Appendix \ref{app:comment_r0_r3}, we briefly comment on $\hat{r}_{t}^{0}$ and $\hat{r}_{t}^{3}$.

Equation (\ref{eq:Lindblad_eq_2}) can be rewritten as the following equation ($\hat{p}$ is the momentum operator), 
\begin{equation} \label{eq:struc}
\partial_t \vec{\hat{{r}}} = \mathbf{P} \hat{p} \vec{\hat{{r}}} + \mathbf{P}^{\dag} \vec{\hat{{r}}} \hat{p} + \mathbf{Q} \vec{\hat{{r}}} \, ,
\end{equation}
on the $4$-component vector (of operators)
\begin{equation}
\vec{\hat{{r}}} \equiv (\hat{r}^0,\hat{r}^1,\hat{r}^2,\hat{r}^3)^{\top} \, ,
\end{equation}
where we have introduced two $4 \times 4$ matrices,
{\small
\begin{equation} \label{eq:PandQ}
\mathbf{P} \equiv \begin{bmatrix}
\cdot & \cdot & \cdot & \mathrm{i} \\
\cdot & \cdot & 1 & \cdot \\
\cdot & -1& \cdot & \cdot \\
\mathrm{i} & \cdot & \cdot & \cdot 
\end{bmatrix} , \  \ \ 
\mathbf{Q}_{\Gamma,m} \equiv \begin{bmatrix}
\cdot & \cdot & \cdot & \cdot \\
\cdot & - \gamma_1 & \cdot & \cdot \\
\cdot & \cdot & -  (\gamma_1 \!  + \! \gamma_2) & - 2 m \\
\cdot & \cdot & 2 m  & - \gamma_2
\end{bmatrix} ,
\end{equation}}
where we have denoted, in the matrices, the zeros by dots to make the writing less cumbersome.
The matrices $\mathbf{P}$ and $\mathbf{Q}_{\Gamma=0,m}$ are anti-Hermitian, because they correspond to the Hamiltonian part of the original equation on $\hat{\rho}$, Eq.\ (\ref{eq:Lindblad_eq_2}), while $\mathbf{Q}_{\Gamma,m=0}$ is Hermitian (more precisely, diagonal and real), and corresponds to the non-Hamiltonian part of the original equation.

\subsubsection{Explicit solution via Fourier transform}

Since we have chosen a vanishing electromagnetic potential, Eq.\ (\ref{eq:vanishing em pot}), a spacetime-independent mass term, Eq.\ (\ref{sent:mass_term}), and a space-independent noise, Eq.\ (\ref{eq:struc}) is diagonal in momentum space. 
We introduce the momentum basis, $\{ \ket{p}, p \in \mathbb{R}\}$. 
Applying $\bra{p}$ on the left of Eq.\ (\ref{eq:struc}), and $\ket{q}$ on its right, we obtain
\begin{equation} \label{eq:mom-space}
\partial_t \vec{\tilde{r}}_{pq} = \mathbf{G}_{pq} \vec{\tilde{r}}_{pq} \, ,
\end{equation}
where
\begin{equation}
\tilde{r}^{\mu}_{pq}\equiv \bra{p} \hat{r}^{\mu} \ket{q} \, ,
\end{equation}
and where we have introduced the following generator of the transport, 
\begin{equation}
\mathbf{G}_{pq} \equiv \begin{bmatrix}
\cdot & \cdot & \cdot & \mathrm{i}(p-q) \\
\cdot & - \gamma_1 & p+q & \cdot \\
\cdot &-(p+q) & -  (\gamma_1 \!  + \! \gamma_2) & - 2 m \\
\mathrm{i}(p-q)& \cdot & 2 m  & - \gamma_2
\end{bmatrix} \, .
\end{equation}
The solution of Eq.\ (\ref{eq:mom-space}) is well-known, and reads (we reintroduce the time label),
\begin{equation}
\vec{\tilde{r}}_{t,pq} = \mathbf{M}_{pq}(t-t_0) \,  \vec{\tilde{r}}_{t_0,pq} \, ,
\end{equation}
where
\begin{equation} \label{eq:exp}
\mathbf{M}_{pq}(t-t_0) \equiv  e^{(t-t_0)\mathbf{G}_{pq}} \, .
\end{equation}

In position space, the solution of our problem can thus be written explicitly as a two-dimensional Fourier transform, namely,
\begin{equation} \label{eq:sol}
r^{\mu}_{t,xy} = \frac{1}{2\pi} \int \! \! \! \!  \int_{\mathbb{R}^2} dp \, dq \, (\mathbf{M}^{\mu}_{\nu})_{pq}(t-t_0) \tilde{r}^{\nu}_{t_0,pq} \, e^{\mathrm{i}(px-qy)} \, ,
\end{equation}
where we sum over $\nu=0,\hspace{-0.5pt}.\hspace{-0.5pt}.\hspace{-0.5pt}.,3$.
Since the matrix $\mathbf{G}_{pq}$ is diagonalizable and can therefore be exponentiated, Eq.\ (\ref{eq:sol}) provides a formal solution to the dynamics of $r^{\mu}_{t,xy}$. However, the diagonal form (and thus the exponential) is very cumbersome in the massive case, so that the explicit solution does not give insight on the phenomena it describes. Of course, one can always compute these integrals numerically and plot all desired observables. In what follows, we have used this expression only in the massless case, to check that it gives the same result as our numerical integration. The matrix exponential can in this case be performed directly using a symbolic mathematics software, without prior diagonalization. In the following sections, we shall get insight on the different regimes of the dynamics, first by viewing the equations directly in position space, just below. 

\subsubsection{System of equations in position space, and remarks} \label{subsubsec:position_space}

Let us rewrite our system of equations, Eq.\ (\ref{eq:struc}), not in momentum space as above in Eq.\ (\ref{eq:mom-space}), but in position space, by applying $\bra{x}$ on the left of Eq.\ (\ref{eq:struc}), and $\ket{x'}$ on its right, which yields
{\small
\begin{subequations} \label{eq:sys1}
\begin{flalign} 
\ \  \partial_t r^0_{xx'} &= (\partial_x + \partial_{x'}) r^3_{xx'}& \\
\ \  \partial_t r^3_{xx'} &= (\partial_x + \partial_{x'}) r^0_{xx'} - \gamma_2 r^3_{xx'} + 2m r^2_{xx'}&
\end{flalign}
\end{subequations}
\vspace{-0.5cm}
\begin{subequations} \label{eq:sys2}
\begin{flalign} 
\ \  \partial_t r^1_{xx'} &= - \mathrm{i}( \partial_{x} - \partial_{x'}) r^2_{xx'} - \gamma_1 r^1_{xx'} & \\
\ \  \partial_t r^2_{xx'} &= \mathrm{i}( \partial_{x} - \partial_{x'}) r^1_{xx'} - (\gamma_1 + \gamma_2) r^2_{xx'} - 2m r^3_{xx'} \,  , & 
\end{flalign}
\end{subequations}
}
with
\begin{align} \label{eq:def}
r^{\mu}_{xx'} \equiv \bra{x} \hat{r}^{\mu} \ket{x'} \, .
\end{align}
One immediatly sees that the mass couples Eqs.\ (\ref{eq:sys1}) to Eqs.\ (\ref{eq:sys2}).
The case of a non-vanishing mass, $m\neq0$, and no noise, $\Gamma= 0$, simply corresponds to standard, Dirac propagation \cite{Park2012, Demikhovskii2010}, and is recalled in Appendix \ref{app:Dirac-propagation}. 

Using the definition provided in Eq.\ (\ref{eq:def}), one can prove that 
\begin{equation} \label{eq:formula}
(\partial_x + \partial_{x'}) r^{\mu}_{xx'} |_{x=x'} = \partial_x R^{\mu}_x \, ,
\end{equation}
where
\begin{equation} \label{eq:defRmu}
R^{\mu}_{x} \equiv {r}^{\mu}_{xx} \, ,
\end{equation}
so that the above system of four equations, (\ref{eq:sys1}) and (\ref{eq:sys2}), considered for $x=x'$, yields
{\small
\begin{subequations} \label{eq:sys12}
\begin{flalign} 
\ \  \partial_t R^0_{x} &= \partial_x R^3_{x}& \\
\ \  \partial_t R^3_{x} &= \partial_x R^0_{x} - \gamma_2 R^3_{x} + 2m R^2_{x}&
\end{flalign}
\end{subequations}
\vspace{-0.5cm}
\begin{subequations} \label{eq:sys22}
\begin{flalign} 
\ \  \partial_t R^1_{x} &= - \mathrm{i}( \partial_{x} - \partial_{x'}) r^2_{xx'}  |_{x=x'}- \gamma_1R^1_{x} & \\
\ \  \partial_t R^2_{x} &= \mathrm{i}( \partial_{x} - \partial_{x'}) r^1_{xx'} |_{x=x'} - (\gamma_1 + \gamma_2) R^2_{x} - 2m R^3_{x} \,  .  \! \! \! \! \! \! \! \! \! &  \! \! \! \! \! \! \! \! \!
\end{flalign}
\end{subequations}
}
Notice that the $R^{\mu}$s are real since the $\hat{r}^{\mu}$s are Hermitian.

Decomposing $r^{\mu}_{xx'}$ in its real and imaginary parts,
\begin{equation}
r^{\mu}_{xx'} \equiv a^{\mu}_{xx'} + \mathrm{i} b^{\mu}_{xx'} \, ,
\end{equation}
and recalling that the $\hat{r}^{\mu}$s are Hermitian, Eqs.\ (\ref{eq:sys22}) can be written as Eqs.\  (\ref{eq:sys23}) below, which show that the reality of  $R^1$ and $R^2$ is consistent with their evolution equations, since the latter only involve real coefficients and unknowns.

Now, notice that in the two first equations above,   (\ref{eq:sys12}), only the densities, i.e., the quantities taken for $x=x'$, are involved.
It also turns out that one can decouple $R^0_x$ from $R^3_x$ in Eqs.\ (\ref{eq:sys12}), by increasing the order of the equations from $1$ to $2$ in time: after a few manipulations, one indeed realizes that $R^0_x$ and $R^3_x$ follow the same, following equation, Eq.\ (\ref{eq:sys13}),
{\small
\begin{flalign} \label{eq:sys13}
 \ \ \partial_t^2 R^{d}_x + \gamma_2 \partial_t R^{d}_x & = \partial_x^2 R^{d}_x + 2 m \partial_x R^2_x \, , \ \ d=0 \, \text{or} \, 3 \, , &
\end{flalign}
\vspace{-0.5cm}
\begin{subequations} \label{eq:sys23}
\begin{flalign} 
\ \  \partial_t R^1_{x} &= 2  \partial_{x} b^2_{xx'} |_{x=x'} - \gamma_1 R^1_{x} & \\
\ \  \partial_t R^2_{x} &= - 2  \partial_{x}  b^1_{xx'} |_{x=x'} - (\gamma_1 + \gamma_2) R^2_{x} - 2m R^3_{x} \,  . & 
\end{flalign}
\end{subequations}
}

\subsection{$m=0, \Gamma\neq0$: a chirality-flip noise on massless Dirac fermions yields the telegraph equation} \label{subsec:telegraph}

If $m=0$, Eqs.\ (\ref{eq:sys13}) and (\ref{eq:sys23})  become
{\small
\begin{flalign} \label{eq:sys14}
 \ \ \partial_t^2 R^{d}_x + \gamma_2 \partial_t R^{d}_x & = \partial_x^2 R^{d}_x  \, , \ \ d=0 \, \text{or} \, 3 \, , &
\end{flalign}
\vspace{-0.5cm}
\begin{subequations} \label{eq:sys24}
\begin{flalign} 
\ \  \partial_t R^1_{x} &= 2  \partial_{x} b^2_{xx'} |_{x=x'} - \gamma_1 R^1_{x} & \\
\ \  \partial_t R^2_{x} &= - 2  \partial_{x}  b^1_{xx'} |_{x=x'} - (\gamma_1 + \gamma_2) R^2_{x} \,  , & 
\end{flalign}
\end{subequations}
}
which, as mentioned early in Sec.\ \ref{subsubsec:position_space}, decouples Eqs.\ (\ref{eq:sys14}) from Eqs.\ (\ref{eq:sys24}).

\subsubsection{Dynamics of the spatial degree of freedom: no quantumness} \label{subsubsec:spatial_dof}

$R^0$ and $R^3$ are, respectively, the probability density and the left-current density.
They code, together, for the diagonal coefficients of the density matrix in the full Hilbert space.
They follow the same \emph{telegraph equation}, Eq.\ \eqref{eq:sys14}, with characteristic speed and diffusion coefficients $c=1$ and
\begin{equation}
D\equiv \frac{1}{\gamma_2} \, ,
\end{equation}
respectively: the chirality-flip noise causes the massless Dirac fermion to diffuse, in addition to its unitary propagating behavior.
Notice that the phase-flip noise, characterized by $\gamma_1$, has \emph{no effect} on the dynamics of $R^0$, nor on that of $R^3$, in this massless case;
this is commented in Appendix \ref{app:phase-change}.

\emph{Consider Eq.\ (\ref{eq:sys14}) alone: because of the vanishing mass, it contains no quantum feature}.
If one can write down a dynamical equation for the density and current density, and needs no quantum amplitude of probability\footnote{So that no purely quantum phenomena can arise from the coherences $r_{xx'}^d$, $d=0,3$.}, this means that the essence of the quantumness of the system, that is, coherence and entanglement, is, if any, limited to the internal off-diagonal space, described by $\hat{r}^{1}$ and $\hat{r}^{2}$, coupled to each other through the system of equations (\ref{eq:sys2}), which is autonomous (i.e., independent of (\ref{eq:sys1})) because $m$ vanishes.
That Eq.\ (\ref{eq:sys14}) contains no quantum feature is to be understood as the fact that the chirality-flip noise affects the spatial dynamics in \emph{a purely classical manner, i.e., not via entanglement}.

In other words, the telegraph equation can be derived from a purely non-quantum modeling of the system.
In particular, if viewed as a continuum limit of some discrete-spacetime dynamics, our system corresponds to a \emph{persistent classical random walk} \cite{Randall2003}.
This is in contrast with the simpler, well-known \emph{classical random walk}, which (i) is the one that is usually considered when the transport has no relativistic feature, and which (ii) leads to diffusion in the continuum\footnote{The \emph{mathematical} connection, via analytical continuation, between the standard, unitary, i.e., non-noisy DTQW, and the telegraph equation, is well-known \cite{Gaveau1984, FUKUSHIMA2017, Arnault17}.
In the present work, the connection is not merely mathematical, but \emph{physical}, via the introduction of noise in the unitary dynamics.
More generally, whether the existence of such a type of connection via analytic continuation implies a physical connection when introducing a noise is an interesting question to be investigated.
} \cite{ROUPHydroLim}.
The telegraph equation can model the propagation of classical waves of light/electricity with dissipation (in wires, for example, hence its name) \cite{Randall2003}.
At short (long) times, propagation (diffusion) dominates over diffusion (propagation) \cite{Randall2003}.
Notice that the telegraph equation was proposed by Cattaneo to model relativistic diffusions and can be viewed as a precursor of Extended Thermodynamics models.


In the light of the comments of the previous paragraph, the massive noisy quantum model of the present work, Eq.\ (\ref{eq:Lindblad_eq_2}), can be seen as a quantum model of relativistic diffusion.
Other models which could be qualified as such for the same reasons, have been considered in the litterature, but apparently mostly with a noise introduced \emph{directly on the spatial d.o.f.} \cite{Cabrera2016, H2009a, H2009b, Falci2010}.
A particularity of the present work is thus to \emph{introduce the noise on the internal d.o.f.\ only.}
The solution of the telegraph equation, Eq.\ (\ref{eq:sys14}), is given by Eq.\ (\ref{eq:tele2}) for $\gamma_1 = 0$, so that $b=0$ and $\kappa = \gamma_2$ in Eqs.\ (\ref{eq:kandb}).

\subsubsection{Dynamics of the internal d.o.f.: standard bit-flip decohering dynamics, \emph{classically} coupled to the spatial degree of freedom}

Without the external d.o.f. (i.e., replace $S(\hat{p})$ by $\mathbb{1}$ in Eq.\ (\ref{eq:evolution_op})), we are left with the two standard error channels \cite{Cappellaro_course, Arsenijevi2017} that we have introduced on the coin (see Eqs.\ (\ref{eq:sys1}) and (\ref{eq:sys2})): the populations' difference $r^3$ decays exponentially with a rate $\gamma_2$, and the real (resp. imaginary) part of the coherences, $r^1$ (resp. $r^2$), also decays exponentially, with a rate  $\gamma_1$ (resp. $\gamma_1 + \gamma_2$).
These two coin error channels\footnote{By definition, a qubit error channel maps a pure state to a mixed state.} are \emph{purely} and \emph{fully} decohering, i.e. (this is our terminology), they make the coherences decrease, as time increases, \emph{monotonically} and \emph{down to zero}, respectively, and in any basis of the internal space; that the populations' difference go to zero is also independent of the basis\footnote{These results can be checked by a simple, direct computation.}.

Let us now reconsider the external d.o.f..
%
%
Similarly to Eq.\ (\ref{eq:formula}), one can prove the following identity,
\begin{equation} \label{eq:formula_minus}
(\partial_x - \partial_{x'}) r^{\mu}_{xx'} |_{x=-x'} = \partial_x T^{\mu}_x \, ,
\end{equation}
where
\begin{equation}
T^{\mu}_x \equiv  r^{\mu}_{x,-x} \, ,
\end{equation}
measures the coherence between the states $\ket{x}$ and $\ket{-x}$,
and inserting Eq.\ (\ref{eq:formula_minus}) for $\mu=1,2$ into Eqs.\ (\ref{eq:sys2}) yields, for $m=0$,
\begin{subequations} \label{eq:sys121}
\begin{flalign} 
\ \  \partial_t T^1 &= - \mathrm{i}\partial_x T^2 - \gamma_1 T^1 & \\
\ \  \partial_t T^2 &= \mathrm{i} \partial_x T^1 - (\gamma_1 + \gamma_2) T^2 \, . &
\end{flalign}
\end{subequations}
Analogously to the manipulation performed to go from Eqs.\ (\ref{eq:sys12}) to Eqs.\ (\ref{eq:sys13}), one can actually decouple $T^1$ and $T^2$: they follow the same telegraph equation as $R^0$ and $R^3$ but with a modified diffusion coefficient,
\begin{equation}
D' \equiv \frac{1}{2 \gamma_1 + \gamma_2} \, ,
\end{equation}
 and an additional self source of decoherence induced by $\gamma_1$,
\begin{align} \label{eq:sys14bis}
\partial_t^2 T^{i} + \kappa  \partial_t T^{i} & = \partial_x^2 T^{i} + b T^i  \, , \ \ i=1 \, \text{or} \, 2 \, ,
\end{align}
with
\begin{subequations} \label{eq:kandb}
\begin{align}
\kappa &\equiv 2 \gamma_1 + \gamma_2 \geq 0 \\
b &\equiv - \gamma_1 (\gamma_1 + \gamma_2) \leq 0 \, .
\end{align}
\end{subequations}
The solution of this equation, provided on page 217 of Ref.\ \cite{book_Polyanin_2002}, reads
\begin{align} \label{eq:tele2}
F_{t,x} &= \frac{1}{2} e^{-\frac{\kappa}{2}t} \left[ f_{x + t} + f_{x-t} \right] \\
&+ \frac{\gamma_2}{2} \frac{t}{2} e^{-\frac{\kappa}{2}t} \int_{x-t}^{x+t} \! \! dy \, \frac{ I_1( \frac{\gamma_2}{2} z_y ) }{z_y} f_{y}  \nonumber\\
&+ \frac{1}{2} e^{-\frac{\kappa}{2}t} \int_{x-t}^{x+t} \! \! dy \, I_0(\tfrac{\gamma_2}{2} z_y) \left[ g_{y} + \frac{\kappa}{2} f_{y} \right] \, , \nonumber
\end{align}
where
\begin{equation}
z_y \equiv (t^2 - (x - y)^2)^{1/2} \, ,
\end{equation}
\begin{equation}
I_{\nu}(X) \equiv \sum_{n=0}^{+\infty} \frac{(X/2)^{\nu + 2n}}{n! \, \Gamma(\nu + n + 1)} \, ,
\end{equation}
is the modified Bessel function of the first kind, and where we need two initial conditions because the equation is of order 2 in time,
\begin{subequations}
\begin{align}
f_x &\equiv F_{0,x} \\
g_x &\equiv \partial_tF|_{0,x} \, .
\end{align}
\end{subequations}

The first thing to mention is that, if both the initial function, $T^i_{t=0}$, and the initial time derivative, $\partial_t T^i|_{t=0} $, both vanish, then $T^i_t = 0$ for any $t$, i.e., \emph{the dynamics generates no coherence between $x$ and $x'$}.
This is a consequence of choosing both a vanishing mass and a purely decohering noise.
If $\gamma_1$ vanishes, no decoherence comes from self sources anymore (phase-flip channel); the remaining decoherence only comes from the chirality-flip channel, and, as already mentioned, the dynamics followed by both $T^1$ and $T^2$ is exactly the same as that followed by $R^0$ and $R^3$, and can be viewed as the consequence of a purely classical coupling between the internal and the external d.o.f.s.
In summary: the initial amount of coherence between the two internal states which is initially introduced in the system, is, as coherence between $x$ and $-x$, spatially transported classically exactly as the probability density.

\begin{figure*}[t]
\hspace{-0cm}
\centering
\includegraphics[width=18cm]{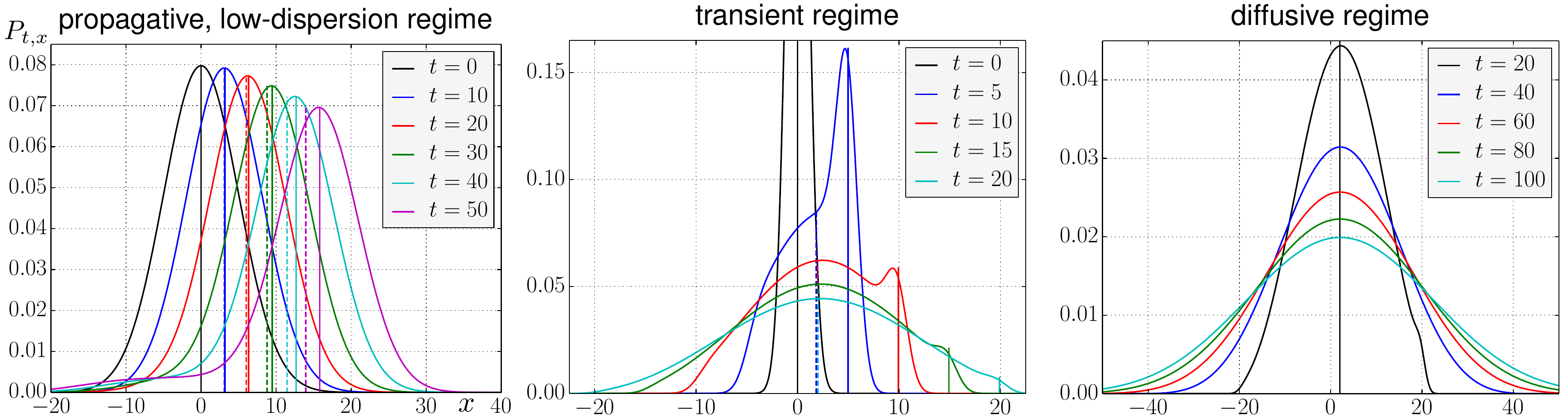}
\caption{Probability density $P_{t,x} \equiv R^0_{t,x}$ (see Eq.\ \eqref{eq:defRmu}) of the massive, relativistic particle experiencing the (quantum) relativistic diffusion governed by the Dirac Lindblad equation \eqref{eq:Lindblad_eq_2} with $\gamma_1=0$, as a function of the position $x$ on the line, for different times $t$.
 The first thing to mention is that the parameters and initial wavefunction -- a (Gaussian) postive-energy wavepacket, Eq.\ \eqref{eq:state_evol}), of the standard Dirac equation -- have been chosen such that the dynamics is essentially classical, i.e., non-quantum, see the explanations in the main text.
The dynamics clearly displays, as time evolves, three regimes.
The first, propagative, low-dispersion regime, from $t=0$ to a $t_1$ defined in Sec.\ \ref{subsubsec:num_stud}, is illustrated in the left panel, for which we have chosen  $\gamma\equiv\gamma_2 = 0.05$, $p_0=1$ and $m=3$, so that  the group velocity $v_g \simeq 0.31$, and $\sigma = 0.1$, so that $\sigma/p_0 = 1/10$, which is why the global propagation dominates over dispersion:
Indeed, one can see that the mean position $\langle x \rangle_t$ of the distribution (see Eq.\ \eqref{eq:meanpos}), represented by dashed vertical lines, differs very little from the ballistic group motion $v_gt$, represented by solid vertical lines;
Of course, the discrepancy between both increases with time.
In the middle and right panels, we have chosen $\gamma= 0.5$,  $p_0=5$ and $m=0.5$, so that  the group velocity $v_g \simeq 0.995$, and $\sigma = 0.5$ (so that $\sigma/p_0 = 1/10$ and we should still be, at least initially, in a low-dispersion case).
Notice that the parameters used in the plots have been chosen in order to qualitatively better display each of the three regimes. The second, transient regime, from $t_1$ to a $t_2$ defined in \ref{subsubsec:num_stud}, is illustrated in the middle figure:
One can very clearly see how diffusion progressively takes over the ballistic motion of the initial density peak.
The third, diffusive regime, from $t_2$ to infinity, is illustrated in the right figure:
One can easily check numerically that this regime tends towards a standard diffusion, with variance $4Dt$, where $D=1/\gamma$. \label{fig:candy}}
\end{figure*}

\begin{figure*}[t]
\hspace{-0cm}
\centering
\includegraphics[width=18cm]{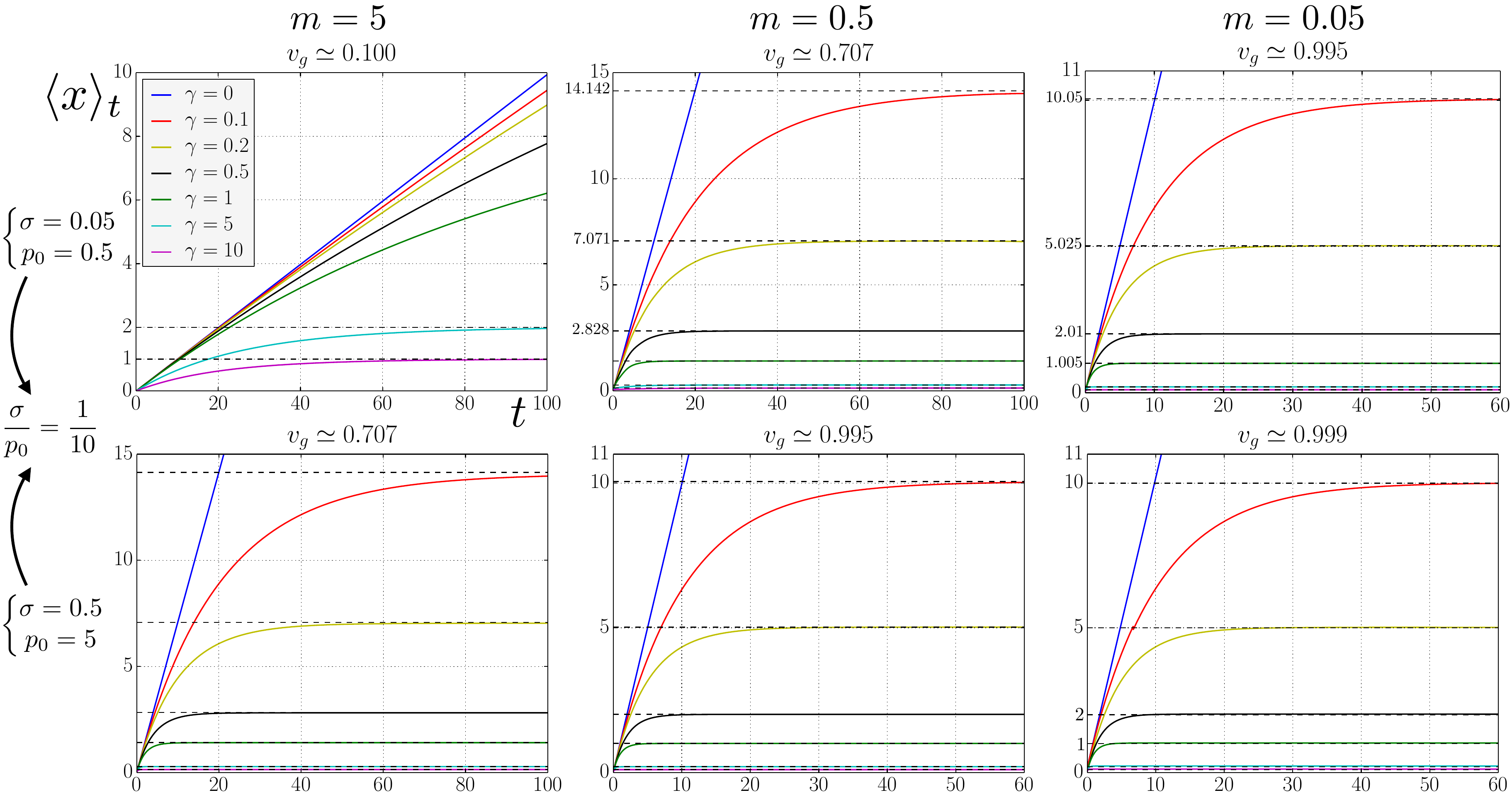}
\caption{Time evolution of the mean position, $\langle x \rangle_t$ (see Eq.\ \eqref{eq:meanpos}), of an initial (Gaussian) positive-energy wavepacket of the standard Dirac equation, see Appendix \ref{subsec:initial}, evolved through the Dirac Lindblad equation\eqref{eq:Lindblad_eq_2} with $\gamma_1=0$ and notation $\gamma_2=\gamma$, i.e., a sole chirality-flip noise.
The first (resp. second) row of plots corresponds to fixing a width $\sigma=0.05$ and a mean momentum $p_0=0.5$ (resp. $\sigma=0.5$ and $p_0=5$), and the ratio $\sigma/p_0=1/10$ is fixed on all plots.
The first column of plots corresponds to $m=5$, the second to $m=0.5$, and the third to $m=0.05$.
We are in the low-dispersion regime, where the width of the initial wavepacket is much smaller than the initial average momentum, $\sigma \ll p_0$, so that the competition is expected to be between the global propagation of the wavepacket and its diffusion due to the noise, the second one increasingly dominating over the first one as time evolves.
The wavepacket initially evolves ballistically at speed $v_g$, but later approaches asymptotically a limit position $x_{\text{lim}}(v_g,\gamma) \equiv 1 / (v_g \gamma) \equiv D/v_g$, see Eq.\ \eqref{eq:xmax}, represented by dashed horizontal lines.
After reaching the limit position, the probability distribution seems to experience an exact diffusion, see Fig.\ \ref{fig:exponent}.
One can see that the top-center (top-right) and bottom-left (bottom-center) figures seem to be almost the same, which suggests that the parameters that characterize the dynamics are (apart from $\gamma$), the ``dispersion/(global propagation)'' ratio,  i.e., $\sigma/p_0$, which characterizes the degree of dispersiveness, and (ii) the group velocity $v_g$, which characterizes the initial global propagation.
The role played by $v_g$, not only up to $x_{\text{lim}}$, but also after, is discussed further in the right-most series of plots of Fig.\ \ref{fig:exponent}.
Notice that, as $v_g$ converges to $1$, $x_{\text{lim}}$ converges to $D$.
 \label{fig:xmax}}
\end{figure*}

\begin{figure*}[t]
\hspace{-0cm}
\centering
\includegraphics[width=18cm]{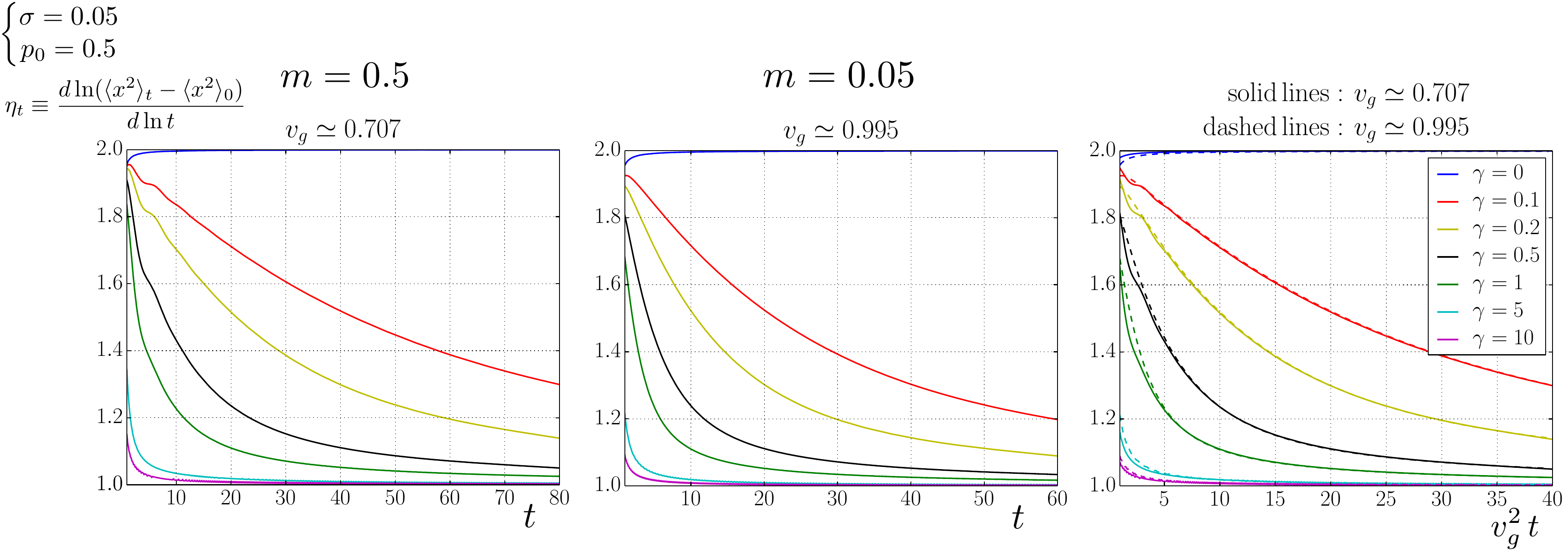}
\caption{Time evolution of the exponent $\eta_t$ (see Eq.\ \eqref{eq:exponent}). The choices made are exactly the same as in the top row of Fig.\ \ref{fig:xmax}, characterized by $\sigma=0.05$ and $p_0=0.5$. 
We know that our numerical code fails for too small times.
In the case $\gamma=0$, the Fourier-transform solution is computationally easy to plot, and we have checked that in this case $\eta_t$ is $2$ whatever $t$.
We believe that $\eta_{t=0}$ should be $2$ (exact ballistic motion) whatever $\gamma$.
It seems from the plot that $\eta_t$ always tends to $1$ when $t$ tends to infinity, i.e., that the dynamics tends towards an exact diffusion.
The left and middle plots manifestly coindice if ploted as a function of $v_g^2 t$.
 \label{fig:exponent}}
\end{figure*}

\subsection{$m\neq0, \Gamma\neq0$: chirality-flip noise on massive Dirac fermions (in the low-dispersion, that is, semi-classical regime)} \label{subsec:complete-case}

In Appendix \ref{app:Dirac-propagation}, we recall the case, $(m\neq 0, \Gamma = 0)$, of standard, Dirac propagation \cite{Park2012, Demikhovskii2010}.
Now, we want to investigate how the (sole) chirality-flip noise influences a \emph{massive} Dirac fermion.
%

Already in the noiseless case, one can distinguish two regimes: (i) a \emph{low-dispersion} regime, in which the global propagation, i.e., the average speed of the distribution, dominates over dispersion, i.e., over the average speed at which the distribution spreads with respect to the mean position, and (ii) its counterpart, the \emph{dispersive} regime.
Let us, for simplicity, focus on the first regime, that is, the non-quantum, or, rather, as we have called it, low-dispersion one, that can be approximately described as the propagation of a classical wave (in vaccum, a classical wave does not disperse)\footnote{We say ``low-dipersion'' rather than ``non-quantum'' because we consider here no potential energy, so that the counterpart, dispersive regime is not ``that quantum'' either, in the following sense. Consider first the non-relativistic regime: it is well-known that, without a potential energy, the Schr{\"o}digner equation can be seen as a classical wave equation in a dispersive medium, with dispersion relation $p^2/2m$. Now, if we are not in the non-relativistic regime, the only quantum feature is Zitterbewegung; apart from it, free Dirac propagation could be described by a scalar amplitude propagating classically in a dispersive medium, with dispersion relation $\sqrt{p^2+m^2}$.}. That is, let us study the classical features of our dynamics, Eq.\ (\ref{eq:sys12}). As induced from the former, massless case, in Sec.\ \ref{subsec:telegraph}, it should be meaningful to qualify this classical dynamics as a \emph{massive relativistic diffusion}.

\subsubsection{Validity of the low-dispersion regime (noiseless study)}
\label{subsubsec:validity}

Let us consider the noiseless case, detailed in Appendix \ref{app:Dirac-propagation}.
If $\sigma$, the momentum spread of the initial (Gaussian) positive-energy wavepacket, Eq.\ (\ref{eq:state_evol}) for $t=0$, is much smaller than the initial average momentum $p_0$, that is, if
\begin{equation} \label{eq:low-disp_cond1}
\frac{\sigma}{p_0} \ll 1 \, ,
\end{equation}
then, intuitively, dispersion should be negligible with respect to propagation during some time.
Let us evaluate this more precisely.
One can prove (not shown) that, if Condition (\ref{eq:low-disp_cond1}) is satisfied, then one can approximate the dispersion relation $E_p = \sqrt{p^2+m^2}$ by a quadratic function of $p-p_0$ by Taylor expanding it around $p_0$, that is, one can make what we call the quadratic-dispersion-relation (QDR) approximation, which, conveniently, enables to do analytical computations (Gaussian integrals).
In the QDR approximation, the mean position and spread are respectively given by 
\begin{equation}
v_g \equiv \left. \frac{\partial E_p}{\partial p} \right|_{p=p_0} = \frac{p_0}{\sqrt{p_0^2 + m^2}}
\end{equation}
and by Eq.\ (\ref{eq:vdisp}).
One can then prove (not shown) by an explicit computation that Condition (\ref{eq:low-disp_cond1}) actually ensures
\begin{equation}
\frac{v_d}{v_g} \ll 1 \, ,
\end{equation}
that is, that we are in the low-dispersion regime.

\subsubsection{Physical quantities to be studied}

Let us introduce the first moment of the probability distribution, i.e., the mean position,
\begin{equation} \label{eq:meanpos}
\langle x \rangle_t \equiv \int_{\mathbb{R}} \! dx \, x P_{t,x} \, ,
\end{equation}
where $P_{t,x} \equiv R^0_{t,x}$ is the presence density, or probability distribution, defined in Eq.\ \eqref{eq:defRmu}.

Let us also introduce the second, \emph{non-centered} moment of the probability distribution, that is,
\begin{equation} \label{eq:second_mom}
\langle x^2 \rangle_t \equiv \int_{\mathbb{R}}  \! dx \, x^2 P_{t,x} \, .
\end{equation}

Finally, let us introduce the exponent,
\begin{equation} \label{eq:exponent}
\eta_t \equiv \frac{d \ln (\langle x^2 \rangle_t - \langle x^2 \rangle_0}{d \ln t} \, ,
\end{equation}
of the numerical fit of $\langle x^2 \rangle_t$, defined in Eq.\ \eqref{eq:second_mom}, by a power law, that is,
\begin{equation}
\langle x^2 \rangle_t \underset{\text{fit}}{\simeq}  \alpha t^{\eta} + \langle x^2 \rangle_0 \, ,
\end{equation}
where $\alpha$ is some constant.

\subsubsection{Numerical study}

\label{subsubsec:num_stud}

Consider an initial (Gaussian) positive-energy wavepacket, Eq.\ (\ref{eq:state_evol}) for $t=0$, satisfying Condition (\ref{eq:low-disp_cond1})\footnote{So that, by extrapolating from the noiseless study of the previous section, \ref{subsubsec:validity}, we expect -- when the Hamiltonian part of the Lindblad equation, Eq.\ (\ref{eq:Lindblad_eq_2}), dominates over the noise part -- to be in the low-dispersion regime (rather than the dispersive one).}, and let it evolve according to Eq.\ \eqref{eq:Lindblad_eq_2} with
\begin{subequations}
\begin{align}
\text{choice:}& \ \gamma_1=0 \, , \\
\text{notation:}& \ \gamma_2=\gamma \, .
\end{align}
\end{subequations}
We have implemented this evolution numerically, via an implicit scheme, described in Appendix \ref{app:Num}. 
The dynamics displayed can be split into two or three regimes, as detailed further down.
The simplest, two-regimes description is the following one: first, a \emph{propagative} regime, in which propagation dominates over diffusion, and  second, a \emph{diffusive} regime, in which diffusion dominates over propagation. 
There are several possible criteria to precisely define the ``domination'' one refers to.
In the three-regimes description, illustrated in Fig.\ \ref{fig:candy}, there is transient regime, as we shall see in more detail.
Let us characterize all the aforementioned regimes by analyzing all figures, \ref{fig:candy}, \ref{fig:xmax} and \ref{fig:exponent}.

Let us focus on Fig.\ \ref{fig:xmax}, displaying $\langle x\rangle_t$.
One can easily check (not shown) that the initial time derivative of $\langle x\rangle_t$ is always $v_g$ to a very good approximation.
It is thus natural to introduce a time $t_1$ such that, up to this time, $\langle x\rangle_t$ is approximable $v_g t$.
Of course, $t_1$ depends on the desired precision of this approximation.
From $0$ to $t_1$, the dynamics is thus not only propagative, but also low-dispersion.
One can then define a transient regime from $t_1$ to some $t_2$ such that for $t \geq t_2$, the mean position $\langle x \rangle_t \simeq \text{constant}$, a constant indeed manifestly reached as it can be seen on Fig.\ \ref{fig:xmax}.
Of course, $t_2$ depends on the desired precision of the constant fit of $\langle x \rangle_t$ for $t \geq t_2$.
The regime from $t_2$ to infinity is a diffusive one, as we shall see below.

This was the three-regimes description.
The interest of the two-regimes description is that we naturally introduce a time $t_{\text{mid}}$ which, unlike $t_1$ and $t_2$, is not arbitrary, i.e., depending on some desired precision, but \emph{characteristic of the dynamics}. 
This $t_{\text{mid}}$ could be, e.g., when the second derivative of $\langle x \rangle_t$ reaches its maximum (see Fig.\ \ref{fig:xmax}), or when it is the second derivative of $\eta_t$ which reaches its maximum (see Fig.\ \ref{fig:exponent}).

Now, that the dynamics always tends towards an exact diffusion for $t\rightarrow +\infty
$, i.e., with a variance scaling as $t$, seems to be expected from Fig.\ \ref{fig:exponent}.
One can easily check (not shown) that the diffusion coefficient is, in all shown cases, that of the massless case, $D=1/\gamma$, i.e., that the variance equals $4Dt$, as expected from the fact that the Lindblad equation is linear and that there is no other diffusion term than that with diffusion coefficient $D$.

Let us comment on the limit position $x_{\text{max}}$ seen on Fig.\ \ref{fig:xmax}.
In all the cases studied in Fig.\ \ref{fig:xmax}, this limit position turns out to be very well approximated by
\begin{equation} \label{eq:xmax}
x_{\text{max}} \underset{\text{fit}}{\simeq}  x_{\text{lim}}(v_g,\gamma) \equiv \frac{1}{v_g\gamma} \equiv \frac{D}{v_g}  \, ,
\end{equation}
whose numerical values have been represented by black, dashed horizontal lines.
Both the dependencies in $v_g$ and $D$ can be understood qualitatively by extrapolating from the massless case, where the probability density $P_{t,x} \equiv R^0_{t,x}$ follows a telegraph equation, Eq.\ \eqref{eq:sys14}, that we rewrite here,
\begin{equation} \label{eq:telegraphBIS}
\frac{D}{c^2} \partial_t^2 P_x + \partial_t P_x = D\partial_x^2 P_x \, .
\end{equation}
We recall that here the characteristic speed is $c=1$.
We have omitted the time index of $P_{t,x}$ as in Eq.\ \eqref{eq:sys14}.

What accounts for $x_{\text{max}}$ diminishing when $D$ does is the following.
As $D$ diminishes, the second time derivative in Eq.\ (\ref{eq:telegraphBIS}), $(D/c^2) \partial_t^2 P_x$, diminishes, while the first time derivative, $\partial_t P_x$, remains unaffected, so that propagation faints with respect to diffusion.
This is a remarkable specificity of the telegraph equation.
Indeed, $D$ is involved not only in the usual diffusion term, $D\partial_x^2 P_x$, but also in the propagation term, $(D/c^2) \partial_t^2 P_x$, and both terms have the same variations as $D$ (they are, more precisely, linear with $D$), so that a smaller $D$ not only implies, via the usual diffusion term, a slower diffusion, \emph{but also, via the propagation term, that the diffusion regime is reached earlier}.
The telegraph equation thus appears as the particular case $d = D$ of the following, more general equation, $(d/c^2)\partial_t^2 f + \partial_t f = D \partial_x^2 f$, in which the diffusion coefficient $D$ of the final diffusive regime is, this time, independent from the time taken to reach this diffusive regime, controled by $d$ (and $c$).

That $x_{\text{max}}$ diminishes when $v_g$ increases could possibly seem counter-intuitive.
The following explanation can actually account for it.
In the telegraph equation, that the characteristic speed $c$ increases makes propagation faint with respect to diffusion, i.e., makes the diffusive regime be reached earlier.
Now, one can argue that  in the massive case, the same mechanism happens, but with a characteristic speed which is not $c$ anymore, but $v_g$.

It is interesting to put this limit-position effect in perspective with an effect predicted in the Stern-Gerlach experiment \cite{Gomis2016}.
In this case, the system also experiences entanglement between the internal and the spatial degrees of freedom. 
However, the noise is assumed to be described by the Caldeira-Leggett model, so that it acts on the spatial part, not the internal one.
One finds a limit momentum, rather than a limit position.

\section{Continuum limit for discrete-time quantum walks with temporal coin noise depending smoothly on the position}
\label{sec:spatiotemp}

\subsection{Adding spatial randomness on top of the temporal randomness of the coin unitary}

\subsubsection{Introduction, and $M$-point function}

We want to allow the temporally random coin unitaries of Sec.\ \ref{subsec:random-unitaries} to be random also spatially.
We thus introduce a random variable $\Omega^{l}_{t,x}$ for each lattice position $x\in \mathbb{L}$, where, for more definiteness, we have considered, instead of $\mathbb{Z}\epsilon$, a \emph{finite} lattice
\begin{equation}
\mathbb{L}\equiv \{ x_1, ..., x_M\} \, ,
\end{equation} 
with $M$ some positive integer.
From now on, we will omit the specification ``$ \in \mathbb{L}$'' when writing ``$x \in\mathbb{L}$''.
%
We keep the temporal independence of the random variables, i.e., $\Omega^{l}_{t_1,x}$ is independent on $\Omega^{l'}_{t_2,x'}$ for $t_2\neq t_1$, whatever $(l,x,l',x')$ (this, in particular, implies the classical Markovianity of the noise).
We assume $l$-independence in space, i.e., $\Omega^{l_1}_{t,x}$ is independent on $\Omega^{l_2}_{t,x'}$ for $l_2\neq l_1$, whatever $(x,x')$.
Now, we do \emph{not} assume spatial independence at fixed $t$ and $l$, i.e., a family of real numbers
\begin{equation}
(\omega_{t,x}^l)_{x} \equiv (\omega_{t,x_1}^l, ..., \omega_{t,x_M}^l) \in \mathbb{R}^M \, ,
\end{equation}
where
\begin{equation}
M \equiv \text{number of sites of the lattice} \, ,
\end{equation}
is issued from a sampling of the family of random variables $(\Omega_{t,x}^l)_{x}$ according to some \emph{arbitrary} $M$-point function (i.e., probability distribution), that we denote by $p^{l,(M)}_{t,(x)_{x}}$\footnote{The temporal independence assumed above simply means being able to give oneself such a $p^{l,(M)}_{t,(x)_{x}}$ at any $t$, without knowing the past history, i.e., the values $(\omega^l_{t',x'})_{\{t' \leq t \}, \{x'\} }$.}, where
\begin{equation}
(x)_x \equiv (x_1,...,x_M) \, .
\end{equation}
The family $(\omega_{t,x}^l)_{x}$ has thus a probabilistic weight $p^{l,(M)}_{t,(x)_{x}}((\omega^l_{t,x})_{x})$\footnote{That this $M$-point function is arbitrary implies, in particular, that it is not necessarily given, as it would be for independent random variables, by a product $p^{l,(M)}_{t,(x)_{x}}((\omega^l_{t,x})_{x}) = \prod_{x}  p^{l,(1)}_t(\omega^l_{t,x})$. Notice that we have not allowed the $1$-point function $p^{l,(1)}_t$ to depend on $x$, i.e., we have impose translational invariance for the noise. Below in the main text, we extend the definition of translational invariance for non-vanishing correlations between the random variables.}; 
in other words, $p^{l,(M)}_{t,(x)_{x}}((\omega^l_{t,x})_{x})$ is the probability (density) of the event ``$\Omega^l_{t,x_1}= \omega^l_{t,x_1}$ and $\Omega^l_{t,x_2}= \omega^l_{t,x_2}$, ..., and $\Omega^l_{t,x_M}= \omega^l_{t,x_M}$''.
As in Sec.\ \ref{subsec:random-unitaries} for the spatially homogeneous noise, we assume the stationarity of the noise: $p^{l,(M)}_{t,(x)_{x}} = p^{l,(M)}_{(x)_{x}}$.

Instead of the evolution of Eq.\ (\ref{eq:evol_rho}), we thus have to consider an evolution of the form
\begin{equation} \label{eq:QW-spatiotemp}
\hat{\rho}_{t+\epsilon} = \int \!  \! d\nu \   \mathcal{V}^{\text{QW}}_{(\epsilon \bar{f}_{t,x} + \omega_{t,x})_{x}} \! (\hat{\rho}_t) \, ,
\end{equation} 
where the integration measure $d\nu$, which  satisfies the normalization condition $\int  \! d\nu =1$, is given by
\begin{equation}
d\nu \equiv (d\omega_t)^M \, P^{(M)}_{(x)_{x}}((\omega_{t,x})_{x}) \, ,
\end{equation}
having introduced the notation
\begin{equation}
(d\omega_t)^M \equiv  \prod_{l=0}^3 \prod_{x} d\omega^l_{t,x} \, ,
\end{equation}
and where, because of the $l$-independence in space, 
\begin{equation}
P^{(M)}_{(x)_{x}}((\omega_{t,x})_{x}) =  \prod_{l=0}^3  p^{l,(M)}_{(x)_{x}}((\omega^l_{t,x})_{x}) \, .
\end{equation}
Since the $\omega_{t,x}$s, varying $x$, are here mute (they are integration variables), and there is no ambiguity about the time $t$ at which we are if we decide not to change, we will use the simplified notation
\begin{equation}
\lambda_x \equiv \omega_{t,x} \, .
\end{equation}
Now, for each family $(\lambda_x^l)_x \in\mathbb{R}^M$, the random-unitaries superoperator is, naturally, given by
\begin{align} \label{eq:one-step-nonun}
&\mathcal{V}^{\text{QW}}_{(\epsilon \bar{f}_{t,x} + \lambda_{x})_{x}} \! (\hat{\rho}_t) \equiv \hat{V}^{\text{QW}}_{(\lambda_{x})_{x}} \ \hat{\rho}_t \ \Big({\hat{V}}^{\text{QW}}_{(\lambda_{x})_{x}}\Big)^{\! \! \dag}\, .
\end{align}
where the random unitaries are, in the present DTQW case,
\begin{equation} \label{eq:random-unit}
\hat{V}^{\text{QW}}_{(\lambda_{x})_{x}} \equiv C_{\epsilon {\bar{f}}_{t,\hat{x}} + {\lambda}_{\hat{x}}} \hat{S} \, ,
\end{equation}
with the position-dependent coin operator
\begin{equation}
C_{\epsilon {\bar{f}}_{t,\hat{x}} + {\lambda}_{\hat{x}}} = \sum_{x} C_{\epsilon {\bar{f}}_{t,x} + {\lambda}_{x}} \ket{x} \! \! \bra{x} \, ,
\end{equation} 
and where, to lighten notations, we have used
\begin{equation}
\hat{S} \equiv S(\hat{p}) \, ,
\end{equation}
given in Eq.\ (\ref{eq:shift}).
\noindent
As in the spatially homogeneous case in Sec.\ \ref{subsec:random-unitaries}, the (densities of) Kraus operators are simply the random unitaries $\hat{V}^{\text{QW}}_{(\lambda_{x})_{x}}$; there is one such operator for each $4M$-uple $(\lambda_{x})_{x}$.

\subsubsection{We only need the $2$-point function because the noise is local}

We qualify a spatiotemporal noise $(\lambda_y)_y$ as local if the random unitary $\hat{V}_{(\lambda_y)_y}$, with a priori arbitrary dependence in ${(\lambda_y)_y}$, has matrix elements of the form
\begin{equation} \label{eq:local-noise}
\bra{x} \hat{V}_{(\lambda_{y})_{y}} \ket{x'} = \sum_{z\in\mathbb{L}} {V}_{\lambda_z}^{xx'} \, ,
\end{equation} 
that is, sums of terms depending, each, on a \emph{single} $\lambda_z$.
It is straigthforward to check that the random unitary $\hat{V}^{\text{QW}}_{(\lambda_y)_y}$, Eq.\ (\ref{eq:random-unit}), is local (with, moreover, the single term $z=x$ in the sum of Eq.\ (\ref{eq:local-noise})).
For a few precisons on local noises, see Appendix \ref{app:local-noise}.

Now, for a local noise as defined in Eq.\ (\ref{eq:local-noise}), the matrix elements $\rho^{xx'}_{t+\epsilon} \equiv \bra{x} \rho_{t+\epsilon} \ket{x'}$ are given by\footnote{To obtain this, just apply $\bra{x}$ (resp.\  $\ket{x}$) on the left (resp.\ right) of Eq.\ (\ref{eq:QW-spatiotemp}) considered, more generally, for a local-noise superoperator $\mathcal{V}$, i.e., made of unitaries satisfying Eq.\ (\ref{eq:local-noise}), not necessarily of the form $\mathcal{V}^{\text{QW}}$.}
{\small
\begin{subequations}
\begin{align} 
\rho^{xx'}_{t+\epsilon}& \equiv \sum_{y,y',z,z'} \int (d\lambda)^M P^{(M)}_{(\tilde{x})_{\tilde{x}}}((\lambda_{\tilde{x}})_{\tilde{x}}) \, V^{xy}_{\lambda_z}  \rho_t^{yy'} {V}_{\lambda_{z'}}^{y'x'}  \\
&=\sum_{y,y',z,z'} \int d\lambda_z d\lambda_{z'} P^{(2)}_{z,z'}(\lambda_z,\lambda_{z'}) \, V^{xy}_{\lambda_z}  \rho_t^{yy'} V_{\lambda_{z'}}^{y'x'} \, .
\label{eq:2point-only}
\end{align}
\end{subequations}}
In going from the first to the second line, we have integrated over the variables that do not appear in the integrand, and assumed that, at any order $n=1,...,M$, there is a single marginal, i.e., not several ones that would be produced by having integrated the higher-order functions over different variables, which is ensured if we assume the $M$-point function to be fully symmetric (i.e., symmetric with respect to all pairs of variables), which, by a natural definition, is a necessary feature of the $M$-point function if \emph{we require the noise to be translationally invariant}; hence, a \emph{sole} $2$-point function appears,
\begin{equation}
P^{(2)}_{z,z'}(\lambda_z,\lambda_{z'}) \equiv \int \bigg( \prod_{r=0}^{R-1} \prod_{ \tilde{x} \neq z,z' }  d\lambda^r_{\tilde{x}} \bigg) P^{(M)}_{(\tilde{x})_{\tilde{x}}}((\lambda_{\tilde{x}})_{\tilde{x}}) \, ,
\end{equation}
where $\lambda_x \equiv (\lambda_x^r)_{r=0,R-1}$, $R \in \mathbb{N}$ being the number of space(time)-dependent parameters which we consider random in space (in the case of the coin operator parametrized by $4$ angles which has been considered in the present work, the maximum $R$ that we can chose is $R=4$, that is, all four angles random in space, and remember that we denoted $r=l$).
The dynamics is completely determined by Eq.\ (\ref{eq:2point-only}), and, hence, by the knowledge of the $2$-point function.
Any $M$-point function, and hence Kraus-operators family $( \hat{V}_{(\lambda_x)_x})_{ (\lambda_x)_x \in \mathbb{R}^{RM}}$, compatible with the $2$-point function characterizing the model, is a valid one to describe that model.
%

\subsubsection{Special form of the $2$-point function for random variables associated to lattice sites, and for a translationally-invariant noise}

By construction of our model, we do not only have a $2$-point function $P^{(2)}_{z,z'}(\lambda,\lambda')$, but we also have that, when $z=z'$, then $\lambda=\lambda'$, so that the $2$-point function must have the form \cite{ACMSWW12}
\begin{align} \label{eq:construction}
&P^{(2)}_{z,{z'}}(\lambda_z,\lambda_{z'}) =  \\ 
& \ \ \ \ \delta_{zz'} P^{(1)}_{z}(\lambda_{z}) \delta(\lambda_z-\lambda_{z'}) + (1 - \delta_{zz'}) P^{(2),\neq}_{z,z'}(\lambda_z,\lambda_{z'}) \, ,
 \nonumber
\end{align}
where  $P^{(1)}_{z}(\lambda_{z})$ is the $1$-point function, and $P^{(2),\neq}_{z,z'}$ is a $2$-point function which need only make sense for $z\neq z'$, i.e., $P^{(2),\neq}_{z,z'}(\lambda_z,\lambda_{z'})$ is, for $z=z'$, an arbitrary and irrelevant $\mathbb{R}^+$-number.

Requiring the noise to be spatial translationally invariant means requiring
{\small
\begin{subequations} \label{eq:trans_inv}
\begin{align}
&\forall (z,\lambda) \in \mathbb{L}\times\mathbb{R}, \ \ P^{(1)}_z(\lambda) = P^{(1)}(\lambda) \label{eq:trans_inv1}  \\
&\forall (z,z',\lambda,\lambda') \in \mathbb{L}^2\times\mathbb{R}^2, \ \ P^{(2)}_{zz'}(\lambda,\lambda') = w_{|z-z'|}(\lambda,\lambda') \label{eq:trans_inv2} \\ 
& \ \ \ \ \ \ \ \ \ \ \ \ \ \ \ \ \ \ \ \ \ \ \ \ \ \ \ \ \ \ \ \ \ \ \ \ \ \ \ \ \ \ \ \ \ \ \ \, = w_{|z-z'|}(\lambda',\lambda) \, ,  \label{eq:trans_inv3}
\end{align}
\end{subequations}}
i.e., (i) that $P^{(1)}(\lambda)$ does not depend on the lattice position $z$, and (ii) that $P^{(2)}(\lambda,\lambda')$ actually depends only on the distance $|z-z'|$, and is an even function of $(\lambda,\lambda')$ (i.e., is symmetric in $(\lambda,\lambda')$).
For random, spatially independent variables, the first condition is of course sufficient, but if we allow non-vanishing $2$-site correlations, the second is also needed.

\subsection{Continuum limit for discrete-time quantum walks with temporal coin noise depending smoothly on the position}
\label{subsec:cont_limite}

\subsubsection{Condition for the temporal continuity of the density operator}

Consider Eq.\ (\ref{eq:2point-only}) for the  DTQW random unitary, Eq.\ (\ref{eq:random-unit}):
{\small
\begin{equation} \label{eq:Tay}
\rho^{xx'}_{t+\epsilon}
= \int \! \! d\lambda_x d\lambda_{x'} P^{(2)}_{x,x'}(\lambda_x,\lambda_{x'}) \ C'_{\lambda_x} \!\bra{x}  \hat{S} \ \hat{\rho}_t  \ \hat{S}^{\dag} \! \ket{x'} C'_{\lambda_{x'}} \, ,
\end{equation}}
where
\begin{equation}
C'_{\lambda_{x}} \equiv C_{\epsilon {\bar{f}}_{t,x} + \tilde{\lambda}_x} \, .
\end{equation}
As in Sec.\ \ref{subsubsection:cont-lim}, we assume
\begin{equation} \label{eq:lambda_scaling}
\Lambda^l_x \equiv  \sqrt{\epsilon} \, \tilde{\Lambda}^l_x \, , \ \ l=0,3, \ x \in \mathbb{L} \, ,
\end{equation}
and change the integration measure in accordance.
The above condition, Eq.\ \eqref{eq:lambda_scaling}, ensures, as in the the case of a sole temporal noise, that $\hat{\rho}_{t+\epsilon}-\hat{\rho}_t$ scales as $\epsilon$, i.e., that $\hat{\rho}$ is a continuous function of time, and hence that $\hat{\rho}$ is approximable by a \emph{differentiable} function of time; it is in this sense that we can write $\partial_t \hat{\rho}$\footnote{In the case where there is no spatial dependence of the parameters of the coin operator, we have also shown the existence of a formal continuum limit by Taylor expanding in that small parameter, $\sqrt{\epsilon}$, \emph{before} making the Kraus integral, so that one may think that it is also only in the above-mentioned sense that we can write $\partial_t \hat{\rho}$. However, one can actually, in this case where there is no spatial dependence, perform the Kraus integral \emph{before} Taylor expanding in $\sqrt{\epsilon}$, and the result actually yields functions which are differentiable in time, so that $\hat{\rho}$ also is, exactly, i.e., does not need to be approximated by a function exhibiting such a feature.}.

\subsubsection{About the difficulties to obtain, once we introduce spatial noise, a PDE description in a sensible continuum limit}
\label{subsubsec:smooth-spatial-dependence}

The question we ask ourselves is whether one can get a sensible limit to the spacetime continuum out of the noisy dynamics described by Eq.\ \eqref{eq:QW-spatiotemp}, i.e., more precisely, whether one can get a PDE for $\hat{\rho}$ in such a limit.
Recall that this is indeed what we have obtained in the case of a purely temporal noise, see Eq.\ \eqref{eq:Lindblad_eq_2}.
Now, because the spacetime-dependent coin-operator parameters are sampled from random variables, $\tilde{\Lambda}^l_x$s, which, for each point $x$ of the 1D spatial lattice, are \emph{different} from one another, then if we take the lattice spacing $\epsilon$ going to zero, the functions of the position resulting from this sampling, i.e., the \emph{realizations} $(\tilde{\lambda}^l_x)_x$ of the spatial noise, will be \emph{discontinuous everywhere on the line}.
Hence, \emph{for each realization} $(\tilde{\lambda}^l_x)_x$ of the spatial noise associated to the evolution $t \rightarrow t+\epsilon$ (i.e., each term in the integral of Eq.\ \eqref{eq:QW-spatiotemp}), $\rho_{t,xx'} \equiv \langle x | \hat{\rho}_t | x' \rangle$ can a priori be considered a continuous function \emph{neither} of $x$ \emph{nor} of $x'$.
Now, at each time step, an average is made over all possible realizations $(\tilde{\lambda}^l_x)_x$ of the spatial noise, see Eq.\ \eqref{eq:QW-spatiotemp}, and it is possible that in certain cases, i.e., with certain constraints, this average \emph{does} only produce continuous, and even differentiable functions $\rho_{t,xx'}$ of $x$ and $x'$.
That being said, this is a delicate topic which would require more work, and we will not treat it in the present article.
Let us simplify the problem and ask ourselves: what are the constraints that one has to impose on the spatial part of the noise for \emph{each} realization $(\tilde{\lambda}^l_x)_x$ of this spatial noise to induce a function $\rho_{t,xx'}$ differentiable in $x$ and $x'$?
A sufficient condition answering this question is the following: such a differentiability of $\rho_{t,xx'}$ as a function of $x$ and $x'$ is trivially guaranteed if we impose all realizations $(\tilde{\lambda}^l_x)_x$ of the spatial noise to be differentiable functions of the position $x$ themselves.
But, imposing this implies that we \emph{{\bfseries loose} the notion of spatial noise in the continuum limit}, that is, in the continuum limit, the superimposed spatial noise introduced at the level of the DTQW, reduces to mere spatial dependence of the temporal noise.
This is the case we are going to treat in the present work.

\subsubsection{Non-explicit Lindbladian form of the continuum limit}

In Appendix \ref{app:spatiotemp}, we show that, if \emph{all} sequences $(\lambda_x^l)_x$ involved in the integral of Eq.\ \eqref{eq:QW-spatiotemp}, correspond, \emph{not} to outcomes of spatially-dependent random variables, but to values taken  by differentiable functions of $x$ (and with which they coincide in the continuum limit), then Eq.\ (\ref{eq:Tay}) admits the following dynamics in the continuum limit, $\epsilon \rightarrow 0$,
{\small
\begin{align} \label{eq:final_spatial}
\partial_t \rho^{xx'} =  -\mathrm{i} \bra{x} [\hat{H}^o, \hat{\rho} ] \ket{x'} +  \mathcal{N}_{\Gamma/2, \kappa_{|x-x'|}}(\rho^{xx'}) \, ,
\end{align} }
where the noise term is
\begin{equation}
\mathcal{N}_{\Gamma/2, \kappa_{|x-x'|}}(\rho^{xx'}) \equiv \sum_{l=0}^2 \gamma_l \mathcal{M}^l_{\kappa^l_{|x-x'|}}(\rho^{xx'}) \, , 
\end{equation}
with
\begin{equation}
\Gamma \equiv (\gamma_0,\gamma_1,\gamma_2) \, ,
\end{equation}
and
\begin{equation}
\mathcal{M}^l_{\kappa^l_{|x-x'|}}(\rho^{xx'})  \equiv  \kappa^l_{|x-x'|} L_l \rho^{xx'} L_l - \frac{1}{2}\{ L_l^{\dag} L_l ,\rho^{xx'} \} \, ,
\end{equation}
where the $L_l$s are given by Eqs.\ (\ref{eq:Lls}) with $L_0 \equiv \mathbf{1}_2$, and where we have introduced the ``variances''
\begin{equation}
\frac{\gamma_l}{2} \equiv \int \! \! d\tilde{\lambda}_y \  \tilde{p}^{l,(1)}(\tilde{\lambda}^l_y) \ (\lambda_y^l)^2 \, ,
\end{equation}
and the ``correlation coefficients''
\begin{equation}
\kappa^l_{|x-x'|} \equiv \frac{c^l_{x,x'}}{\gamma^l / 2} \in [-1,1] \, ,
\end{equation}
where the $c^l_{x,x'}$s are the $2$-point ``correlation functions'',
\begin{equation}
c^l_{x,x'} \equiv \int \! \! d\tilde{\lambda}_x d\tilde{\lambda}_{x'} \ \tilde{p}_{x,x'}^{l,(2)}(\tilde{\lambda}^l_x, \tilde{\lambda}^l_{x'})  \ \tilde{\lambda}^l_x \tilde{\lambda}^l_{x'} \, ,
\end{equation}
which actually depend on $|x-x'|$ only, because of Condition (\ref{eq:trans_inv2}) for the translational invariance of the noise.
Notice that, while the contribution of the noise $l=0$ to the continuum limit was vanishing in Sec.\ \ref{subsubsection:cont-lim}, here it does not, because of the spatial inhomogeneity. Indeed, it is because in general $\kappa^0_{|x-x'|} \neq 1$, that the contibution $l=0$ does not vanish:
\begin{equation}
\mathcal{M}^0_{\kappa^0_{|x-x'|}}(\rho^{xx'})  = ( \kappa^0_{|x-x'|} - 1)\rho^{xx'} \, .
\end{equation}

One can check that Eq.\ (\ref{eq:final_spatial}) is trace preserving, by taking it at $x=x'$, and summing over all $x$s and over $L,R$.
The left-hand side then becomes
\begin{align} \label{eq:left}
\int_{\mathbb{R}} dx \sum_{L,R} \partial_t  [({\rho}^{uu})^{xx}] = \partial_t ( \text{Tr} \hat{\rho} ) \, ,
\end{align}
while the right-hand side is (since the Hamiltonian part is trace preserving)
\begin{equation} \label{right-hand}
 \int_{\mathbb{R}} dx  \left[ ({\rho}^{LL})^{xx} + ({\rho}^{RR})^{xx} \right] \sum_{l=0}^2 \gamma_l \left( \kappa^l_{|x-x'|=0}  - 1\right)  \, .
\end{equation}
By construction of our 2-point function, see Eq.\ (\ref{eq:construction}),
\begin{equation} \label{eq:equal1}
\kappa^l_{|x-x'|=0} = 1 \, , \ \ \ \ \ l=0,...,2 \, .
\end{equation}
The right-hand side, (\ref{right-hand}), thus vanishes, and hence so does the left-hand side, which yields trace preservation.

Notice that the fact that this continuum limit only makes sense for a sole temporal coin noise with smooth spatial variations, and is \emph{not} valid for a superimposed spatial coin noise, implies that the ``correlation coefficient'' $\kappa^l_{|x-x'|}$ is a differentiable function of $|x-x'|$, which in turn is consistent with the fact that our resulting PDE involves, in the Hamiltonian part, derivatives of $\rho_t$ with respect to  $x$ and $x'$.

\subsubsection{Explicit Lindbladian form}

We are going to show that one can derive, from a certain, quite general family of random unitaries, a continuum Lindbladian limit.
Consider the dynamical map ensuing from arbitrary temporal-noise random unitaries $\hat{Q}_{\phi}(\sqrt{\epsilon})$, (i) depending on an arbitrary sequence $\phi\equiv (\lambda_{x})_x$ of values taken by a differentiable function of $x$, with which the sequence coincides in the continuum limit $\epsilon \rightarrow 0$, and (ii) being a function of the square root $\sqrt{\epsilon}$ of the spatiotemporal-lattice spacing $\epsilon$,
\begin{equation} \label{eq:dynamical-map}
\hat{\rho}_{t+\epsilon} = \int \! \! d\nu \, \hat{Q}_{\phi}(\sqrt{\epsilon}) \, \hat{\rho}_t \, \hat{Q}^{\dag}_{\phi}(\sqrt{\epsilon}) \, ,
\end{equation}
where the integration measure satisfies $\int \! d\nu = 1$.
 Assume that the random unitaries have the following Taylor expansion,
\begin{equation} \label{eq:TAYLOR2}
\hat{Q}_{\phi}(\sqrt{\epsilon}) = \mathbb{1} + \sqrt{\epsilon} \hat{Q}^{(1/2)}_{\phi} + \epsilon \hat{Q}^{(1)}_{\phi}  + O(\epsilon^{3/2}) \, .
\end{equation}
Equation (\ref{eq:dynamical-map}) then reads
{\small
\begin{align}
&\hat{\rho} + \epsilon \partial_t \hat{\rho} = \hat{\rho} + \sqrt{\epsilon} \,  \mathscr{H}\big(  \overline{\hat{Q}}^{\raisebox{-1.mm}{\scriptsize$(1/2)$}}   \hat{\rho}_t \big)  \\ \nonumber
& \ \ \ \ +  \epsilon \, \bigg[ \mathscr{H}\big( \overline{\hat{Q}}^{\raisebox{-1.mm}{\scriptsize$(1)$}}  \hat{\rho}_t \big) + \int \! \! d\nu \, \hat{Q}^{(1/2)}_{\phi}   \hat{\rho}_t   \big(\hat{Q}^{(1/2)}_{\phi}\big)^{\!\dag}  \bigg] + O(\epsilon^{3/2}) \, ,
\end{align}}
having introduced the mean value of an operator $\hat{O}_{\phi}$,
\begin{equation}
\overline{\hat{O}} \equiv \int \! \! d\nu \, \hat{O}_{\phi} \, ,
\end{equation}
and the Hermitian-symmetric part of an operator $\hat{A}$,
\begin{equation}
\mathscr{H}(\hat{A}) \equiv  \hat{A} + \hat{A}^{\dag} \, .
\end{equation}
For this expansion, Eq.\ (\ref{eq:TAYLOR2}), to make sense whatever $\epsilon \geq 0$, one needs
\begin{equation}
\mathscr{H}\big(  \overline{\hat{Q}}^{\raisebox{-1.mm}{\scriptsize$(1/2)$}}  \hat{\rho}_t \big) = 0 \, ,
\end{equation}
for which it is sufficient that
\begin{equation}
\overline{\hat{Q}}^{\raisebox{-1.mm}{\scriptsize$(1/2)$}} \equiv \int \! \! d\nu \, \hat{Q}^{(1/2)}_{\phi}  = 0 \, ,
\end{equation}
and one then obtains the following PDE,
{\small
\begin{equation} \label{eq:prelim}
\partial_t \hat{\rho} = \overline{\hat{Q}}^{\raisebox{-1.mm}{\scriptsize$(1)$}}  \hat{\rho}_t + \hat{\rho}_t \big(\overline{\hat{Q}}^{\raisebox{-1.mm}{\scriptsize$(1)$}} \big)^{\!\dag} + \int \! \! d\nu \, \hat{Q}^{(1/2)}_{\phi}   \hat{\rho}_t   \big(\hat{Q}^{(1/2)}_{\phi}\big)^{\!\dag} \, .
\end{equation}}
Now, $\overline{\hat{Q}}^{\raisebox{-1.mm}{\scriptsize$(1)$}} $ is in general not Hermitian, and can be decomposed into a Hermitian and an anti-Hermitian part,
\begin{equation}
\overline{\hat{Q}}^{\raisebox{-1.mm}{\scriptsize$(1)$}}  = \hat{G} +  (-\mathrm{i} \hat{H}) \, ,
\end{equation}
where
\begin{subequations}
\begin{align}
\hat{H} &\equiv \frac{\mathrm{i}}{2} \Big[ \, \overline{\hat{Q}}^{\raisebox{-1.mm}{\scriptsize$(1)$}}  - \big(\overline{\hat{Q}}^{\raisebox{-1.mm}{\scriptsize$(1)$}} \big)^{\!\dag} \Big] \\
\hat{G} &\equiv \frac{\mathrm{1}}{2} \Big[ \, \overline{\hat{Q}}^{\raisebox{-1.mm}{\scriptsize$(1)$}} + \big( \overline{\hat{Q}}^{\raisebox{-1.mm}{\scriptsize$(1)$}} \big)^{\!\dag} \Big] \, ,
\end{align}
\end{subequations}
are both Hermitian.
Equation (\ref{eq:prelim}) can then be rewritten as
\begin{equation} \label{eq:prelim2}
\partial_t \hat{\rho} = -\mathrm{i} [\hat{H},\hat{\rho}] + \{\hat{G}, \hat{\rho} \} + \int \! \! d\nu \, \hat{Q}^{(1/2)}_{\phi}   \hat{\rho}_t   \big(\hat{Q}^{(1/2)}_{\phi}\big)^{\!\dag} \, ,
\end{equation}
where $[\cdot,\cdot{}]$ is the commutator, and $\{\cdot,\cdot{}\}$ the anticommutator.
Notice from this equation that $\hat{H}$ is a Hamiltonian.
Now, requiring that Evolution (\ref{eq:dynamical-map}) be trace preserving implies the following normalization condition,
\begin{equation}
\int \! \! d\nu \, \hat{Q}^{\dag}_{\phi}(\sqrt{\epsilon}) \hat{Q}_{\phi}(\sqrt{\epsilon}) = 1 \, , 
\end{equation}
which, using the Taylor expansion of Eq.\ (\ref{eq:TAYLOR2}), imposes
\begin{equation}
\hat{G} = - \frac{1}{2} \int \! \! d\nu \, \big(\hat{Q}^{(1/2)}_{\phi}\big)^{\!\dag} \hat{Q}^{(1/2)}_{\phi}   \, .
\end{equation}
Plugging this expression of $\hat{G}$ into Eq.\ (\ref{eq:prelim2}) finally yields
\begin{equation}
\partial_t \hat{\rho} = -\mathrm{i} [\hat{H},\hat{\rho}] +  \int \! \! d\nu \bigg[\hat{L}_{\phi}   \hat{\rho}_t   \hat{L}_{\phi}^{\!\dag} -  \frac{1}{2} \left\{ \hat{L}_{\phi}^{\!\dag} \hat{L}_{\phi}, \hat{\rho} \right\} \bigg] \, ,
\end{equation}
which is a Lindblad equation, with Lindblad operators
\begin{equation}
\hat{L}_{\phi} \equiv \hat{Q}^{(1/2)}_{\phi} \, .
\end{equation}
One can apply this general result to recover (i) that of Sec.\ \ref{subsubsection:cont-lim}, with a pure temporal coin noise, and (ii) that of Sec.\ \ref{subsec:cont_limite}, with a temporal coin noise which depends smoothly on the position. \\ \\

\section{Conclusion}

As we discussed in the Introduction, the search for a correct description of diffusive dynamics in relativistic quantum systems has faced historically many difficulties, in the attempt to preserve essential features such as relativistic covariance or causality.
In the non-quantum case, these difficulties have been overcome.
In the quantum case, they are still under study.
In the present paper, we do not address covariance.
We present a model that can be used to simulate some features observed in more involved systems.

Our starting point, see Sec. \ref{sec:basics}, is a DTQW on a one-dimensional lattice, whose walker is subject, see Sec.\ \ref{sec:noise}, to noise \emph{acting on its internal, coin (or chirality) degree of freedom}, that makes it decohere.
We consider two such models of decoherent DTQW.
First, a model with both a coin-flip and a phase-flip channel.
Second, a model of random coin unitary operators (so-called random coin unitaries).
Noise acting on a two-level quantum system (such as the chirality part of a chiral system), appears in many physical scenarios, and is commonly described, microscopically, by spin-boson models\footnote{Reference \cite{Arsenijevi2017} provides such a microscopic model in a framework which is very close to that of the present work; The only difference regarding the noise aspect is that they consider the depolarizing channel, which is an equally weighted sum of the three standard error channels for two-level systems, while we consider only two of these three channels, and with arbitrary weights.}.
Such scenarios include the description of matter in a quantized radiation field, the motion of light particles in metals, or superconducting qubits which are coupled to propagating photons. 
 
Given the update rules that govern the dynamics of DTQWs, their causality is guaranteed by construction.
In fact, an important property of noiseless DTQWs is the ability to reproduce the dynamics of relativistic particles in the continuum limit, i.e., when both the lattice spacing and the time step go to zero.
This also requires that the parameters of the coin operator that controls the dynamics follow this scaling in an appropriate manner.
One can naturally ask the question of what is the continuum limit (if any) of the above decoherent-DTQW models.
As expected, the existence of such a limit also imposes conditions on the behavior of the parameters that characterize the noise, as we approach the continuum.
Within this assumption, we obtain that the two decoherent-DTQW models introduced above admit a common formal continuum limit, namely, a Lindblad equation with a Dirac-fermion Hamiltonian part and, as Lindblad jumps, a chirality flip and a chirality-dependent phase flip, which are two of the three standard error channels for a two-level quantum system.
This, as we may call it, Dirac Lindblad equation, provides a model of quantum relativistic spatial diffusion, which is evidenced both analytically and numerically in Sec.\ \ref{sec:QRD}.
The presence of the chirality, along with its entanglement with the spatial motion, is of course, in our noise model, a crucial ingredient in obtaining such a quantum relativistic system with spatial diffusion, given that the noise acts on the chirality.

We have investigated the resulting dynamics.
For a particle with vanishing mass, the model reduces to the well-known telegraph equation, which yields propagation at short times, diffusion at long times, and exhibits no quantumness, in the sense that it can be described by a wave equation \emph{on the density of presence of the particle}.
On the other hand, the massive case has been analyzed numerically, and exhibits a rich phenomenology. 
We analyzed in detail the dynamics that appears when the initial state is Gaussian, and identified the relevant parameters of the problem.
In the low-dispersion regime, corresponding to an initial momentum width which is much smaller than the initial average momentum, the average position first propagates ballistically, with a velocity that equals the group velocity and, after a transient regime, asymptotically approaches a limit position.

We also extended, in Sec.\ \ref{sec:spatiotemp}, our formal-continuum-limit procedure to temporal noises which depend smoothly on position.
We stress that this does not correspond to adding a spatial noise in any way.

Noiseless quantum walks have numerous applications. 
In quantum algorithmics, they are known to be universal computational primitives. 
In quantum simulation, they can emulate high-energy phenomena like particles propagating in external gauge fields (including a gravitational potential).
And it has been suggested that quantum walks can also serve as building blocks for discrete models of gauge theories.
Thus, stochastic quantum walks such as those considered in this article are useful tools to investigate the effects of decoherence, both in quantum algorithmics
and in quantum simulation.
To be fair, the results presented above are only partial and should be extended, not only to quantum walks on graphs,  but also to many-particle quantum walks, i.e. quantum cellular automata. 
But they reveal, through quantum walks, a profound and unexpected link between quantum algorithms running in a non ideal, `open' quantum computer and relativistic 
diffusions.
The link between decohering gauge theories and relativistic diffusions is perhaps less surprising, but it seems never to have been mentioned in the literature so far.
Speaking quite generally, the results presented in this article show that the vast body of knowledge accumulated on classical relativistic diffusions can contribute to our
understanding of open quantum systems, at least in situations where quantum walks play a natural role. And, vice-versa, studying decoherence of systems which can be modelled with quantum walks can teach us about relativistic diffusions.


\begin{center}

{\centering \bfseries{AKNOWLEDGEMENTS}}

\end{center}

We acknowledge fruitful discussions with José Mar{\'i}a Ib{\'a}{\~n}ez and Isabel Cordero Carri{\'o}n.
P.A. thanks Alba Mar{\'i}a Tortosa Benito for her support, and Justin Gabriel for his friendship.
This work has been funded by the  CNRS PEPs Spain-France PIC2017FR6, the Spanish FEDER/MCIyU-AEI grant FPA2017-84543-P, SEV-2014-0398 and Generalitat Valenciana grant PROMETEO/2019/087.
We also acknowledge support from CSIC Research Platform PTI-001.


%

\appendix

\section{Euler-angles parametrization of $\mathrm{SU}(2)$} \label{app:Euler_angles}

Consider the coin operator of Eq.\ (\ref{eq:chi}) with, for simplicity, spacetime-independent entries. Up to the global phase $\xi^0$, this arbitrary $2\times 2$ unitary matrix,
\begin{equation}
C_{(\xi^0,\xi^1,\theta,\chi)} \equiv  e^{i\xi^0} R_{(\xi^1,\theta,\chi)} \in \mathrm{U}(2) \, , 
\end{equation}
is nothing but an arbitrary coin rotation, which can be written as
\begin{equation} \label{eq:R}
R_{(\xi^1,\theta,\chi)} \equiv 
e^{\mathrm{i}\frac{\xi^1+\chi}{2} \sigma^3} e^{\mathrm{i}\theta \sigma^1}  e^{\mathrm{i}\frac{\xi^1-\chi}{2}\sigma^3} \in \mathrm{SU}(2) \, .
\end{equation}
We have put the dependence on the angles between round brackets to indicate that these angles are constant in spacetime, i.e., only correspond, each, to one real variable.

We have chosen to parametrize this coin rotation with the angles  $\xi^1$, $\theta$, and $\chi$, which are the following linear combinations of the Euler angles of $\mathrm{SO}(3)$ for a passive rotation:
\begin{subequations} \label{eq:linear_comb}
\begin{align}
\psi &= \xi^1 - \chi \\
\phi &=\xi^1+\chi \\
\Theta &= 2\theta \, ,
\end{align}
\end{subequations}
see Ref.\ \cite{Arnault17}, Appendix F.
Notice that $\theta$ is just half of the Euler angle giving the latitude.
Notice also on Eq.\ (\ref{eq:R}) that $\chi$ simply corresponds to a change of coin basis in the equatorial plane of the Bloch sphere.
This parametrization is a compromise between (i) good visualization of the action of the coin rotation on the Bloch sphere, Eq.\ (\ref{eq:R}), which is why we use \emph{almost} the Euler angles -- the only subtlety being, as the reader may have noticed, the visualization of $\xi^1$ \cite{Arnault17} --, and (ii) compactness of writing in a single-matrix form, Eq.\ (\ref{eq:coin_op}), which is why we do the above linear combinations\footnote{Indeed, if we had stuck strictly to using the Euler angles, we would have linear combinations of $\psi$ and $\phi$ in the argument of the exponentials that appear in the matrix.}, Eqs.\ (\ref{eq:linear_comb}).

\section{Quantum continuity equation}
\label{app:comment_r0_r3}

Equation (\ref{eq:Tr}) shows, in particular, (i) that\footnote{The four $\hat{\rho}_t^{uv}$s, $u,v \in \{ L,R \}$, are the components of $\hat{\rho}_t$ on this basis of the mixed coin states that we call canonical, which is induced by the $LR$ basis of the coin pure states: $\hat{\rho} \equiv \sum_{u,v=L,R} \hat{\rho}_t^{uv} \ket{u} \! \! \bra{v}$.}
\begin{equation}
\hat{r}_t^{0} = \text{Tr}_{\text{c}}(\hat{\rho}_t) =  \hat{\rho}_t^{LL} + \hat{\rho}_t^{RR} \, ,
\end{equation}
is the operator corresponding to the probability of presence regardless of the internal state (partial trace of $\hat{\rho}_t$ over the internal d.o.f.), and (ii) that,
\begin{equation} \label{eq:left_current}
\hat{r}_t^{3} = \hat{\rho}_t^{LL} - \hat{\rho}_t^{RR} \, ,
\end{equation}
is the left-current operator. 
Points (i) and (ii) can be illustrated by the fact that one of the four coupled  PDEs implied by Eq.\ (\ref{eq:Lindblad_eq_2}) on the $\hat{r}_t^{\mu}$s, is (we omit the time label to lighten notations),
\begin{equation} \label{eq:continuity_eq_q}
\partial_t \hat{r}^0 = i[\hat{p},\hat{r}^3] \, ,
\end{equation}
which is the quantum-operator version of the 1D continuity equation;
indeed, in position space, i.e., applying $\bra{x}$ on its left, and $\ket{y}$ on its right, Eq.\ (\ref{eq:continuity_eq_q}) delivers
\begin{equation} \label{eq:interm}
\partial_t {r}^0_{xy} = (\partial_x + \partial_{y}) {r}^3_{xy} \, ,
\end{equation}
where ${r}^{\mu}_{xy} \equiv \bra{x} \hat{r}^{\mu} \ket{y}$,
and considering this equation, (\ref{eq:interm}), for $x=y$, yields (see Eq.\ (\ref{eq:formula})),
\begin{equation} \label{eq:continuity_eq}
\partial_t {R}^0_{x} - \partial_x {R}^3_{x} = 0 \, ,
\end{equation}
where $R^{\mu}_{x} \equiv {r}^{\mu}_{xx}$.
Equation (\ref{eq:continuity_eq}) is a standard continuity equation, not specific to a quantum setting.
${R}^0_{x}$ is the probability density, and  $ {R}^3_{x} $ the left-current density.

Notice the following.
``How much'' quantum information contained in Eq.\ (\ref{eq:continuity_eq_q}) manifest itself (so that the equation cannot be reduced to its non-quantum version, Eq.\ (\ref{eq:continuity_eq})), is conditioned to ``how much'' the evolution equation for $\hat{r}^{3}$ contains quantum information as well.
In the present case, we will see in Sec.\ \ref{subsubsec:spatial_dof} see that this demands that the mass $m$ does not vanish.
Now, we know a priori that this condition is, although necessary, not sufficient, because in the non-relativistic limit, the internal and external d.o.f.s do not get entangled by the free dynamics, as mentioned early in Sec.\ \ref{subsubsec:presentation}.
So, another necessary condition for Eq.\ (\ref{eq:continuity_eq_q}) to contain purely quantum information is, for the present model, that the latter is relativistic.

\section{$m\neq0, \Gamma=0$: standard, massive Dirac fermions, a very brief reminder} \label{app:Dirac-propagation}

\subsection{General solution for a pure initial state}

When $\Gamma=0$, the dynamics is unitary. 
Hence, if we do not wish to evolve mixed initial states, the density-operator formalism is not necessary; we can work with the state-vector formalism.
The evolution of an arbitrary pure state  from time $t_0 = 0$ to time $t$ is given by
\begin{equation}
\Psi_{t,x} = \frac{1}{\sqrt{2\pi}} \int_{\mathbb{R}} dp \tilde{\Psi}_{t,p} e^{\mathrm{i}px} \, ,
\end{equation}
where
\begin{equation} \label{eq:decompos}
 \tilde{\Psi}_{t,p} = \alpha^+_p V^+_p e^{-\mathrm{i} E_p t} +  \alpha^-_p V^-_p e^{\mathrm{i} E_p t} \, ,
\end{equation}
is the decomposition, at any time $t$, of the momentum amplitude distribution on the chirality eigenbasis of the considered Hamiltonian, \begin{equation}
h(p) \equiv H^{\text{Dirac}}_{m,0}(p) = \begin{bmatrix}
-p & m \\ m & p
\end{bmatrix} \, ,
\end{equation}
see Eq.\ (\ref{eq:Ho}).
The eigenvalues are
 $\pm E_p$, with
\begin{equation}
E_p \equiv \sqrt{p^2+m^2} \, ,
\end{equation}
and two possible eigenvectors are\footnote{These are non-normalized in the internal space; there is no need of doing so.}
\begin{equation}
V^{\pm}_p \equiv 
\left(
\begin{matrix}
1 \\ \frac{\pm E_p + p}{m}
\end{matrix}
\right) \, .
\end{equation}
The coefficients $\alpha^{\pm}$ of the decomposition, Eq.\ (\ref{eq:decompos}), are, apart from the normalization condition,
\begin{equation} \label{eq:normalization1}
\int_{\mathbb{R}} dp  \left( |\alpha^{+}_p|^2 |\!| V^{+}_p |\!|^2 + |\alpha^{-}_p|^2 |\!|V^{-}_p |\!|^2 \right) = 1 \, ,
\end{equation}
arbitrary complex-valued functions of $p$.

\subsection{Choice of the initial condition}
\label{subsec:initial}

Unless otherwise mentioned, we choose a positive-energy initial state with Gaussian momentum distribution of center $p_0$ and spread $\sigma$, that is,
\begin{subequations} \label{eq:condini}
\begin{align}
\alpha^+_p &= \beta_{p-p_0} \equiv \sqrt{N} \sqrt{g^{\sigma}_{p-p_0}} \\
\alpha^-_p &= 0 \, ,
\end{align}
\end{subequations}
where
\begin{equation}
g^{\sigma}_p \equiv \frac{1}{\sqrt{2\pi}\sigma} e^{-\frac{p^2}{2\sigma^2}} \, ,
\end{equation}
and $N$ is a normalization factor, such that (see Eq.\ (\ref{eq:normalization1}))
\begin{equation}
\int_{\mathbb{R}} dp \,  N  g^{\sigma}_{p-p_0} \left[ 1 + \left( \frac{E_p + p}{m} \right)^2 \right]  = 1   \, .
\end{equation}

\subsection{Group and dispersion velocities}

With the initial condition of Eq.\ (\ref{eq:condini}), the state evolution is
\begin{equation} \label{eq:state_evol}
\Psi_{t,x} = \frac{1}{\sqrt{2\pi}} \int_{\mathbb{R}} dp \, A_{p-p_0} e^{\mathrm{i}(px-E_p t)} \, ,
\end{equation}
where
\begin{equation}
A_{p-p_0} \equiv \beta_{p-p_0} V^+_p \, .
\end{equation}

One can show that a sufficient condition for the dispersion relation, $E_p$, to be considered quadratical, i.e., for what we call the quadratical-dispersion-relation (QDR) approximation to be made, is
\begin{equation}
\sigma \ll p_0 \, ,
\end{equation}
which corresponds to $A_{p-p_0}$ sharply peaked around $p=p_0$.
Within the QDR approximation, the mean position and the centered spread are well approximated by the following, ballistic formulae \cite{Schwartz_course}
\begin{subequations}
\begin{align}
X_t &\equiv \langle x \rangle_t \simeq v_g t  \\
\Sigma_t &\equiv \sqrt{\langle x^2 \rangle_t -  \langle x \rangle^2_t} \simeq \Sigma_0 \sqrt{1 + \frac{v_d^2}{\Sigma_0^2}t^2} \, ,
\end{align}
\end{subequations}
with
\begin{equation} \label{eq:init_Sigma}
\Sigma_0 \equiv \frac{1}{4\sigma^2} + a \, ,
\end{equation}
and where the group and dispersion velocities are respectively given by
\begin{subequations} \label{eq:vgroup_and_vdisp}
\begin{align}
v_g &\equiv \left. \frac{\partial E_p}{\partial p} \right|_{p=p_0} = \frac{p_0}{\sqrt{p_0^2 + m^2}} \\
v_d &\equiv \Gamma|_{p=p_0} \sigma^2 + b \, , \label{eq:vdisp}
\end{align}
\end{subequations}
where
\begin{equation}
\Gamma|_{p=p_0} \equiv \left. \frac{\partial^2 E_p}{\partial^2 p} \right|_{p=p_0} \, .
\end{equation}
The parameters $a$ and $b$, intervening respectively in Eqs.\ \eqref{eq:init_Sigma} and \eqref{eq:vdisp}, are a real numbers that can be computed analytically (within the QDR approximation).

\section{Phase-change noise channel on a massless discrete-time quantum walk} \label{app:phase-change}

In Sec.\ \ref{subsubsec:spatial_dof}, we have mentioned that the phase-flip noise, characterized by $\gamma_1$, has \emph{no effect} on the dynamics of $R^0$ and $R^3$ in the massless case.
This would actually be the case of any \emph{phase-change jump operator}, i.e., with the following properties, (i) it acts solely on the chirality, internal Hilbert space, (ii) it is diagonal, and (iii), it is unitary, that is, it has the form $J=\text{diag}(\text{exp}(i\varphi_L),\text{exp}(i\varphi_R))$, which adds, to each of the two wavefunction components (left- and right-moving), an arbitrary phase ($\varphi_L$ and  $\varphi_R$).
Indeed, the phases of these two wavefunction components do simply not influence the spatial dynamics in the massless case.

Now, if we reintroduce a non-vanishing spacetime-lattice spacing $\epsilon$, i.e., if the walker does experience the discreteness of  spacetime, then any such $J$, e.g., even the identity $\text{diag}(1,1)$, \emph{does} modify the spatial dynamics, but this is simply because applying the noise term $\pi_1 J \hat{\rho}_t J$ in Eq.\ \eqref{eq:depol_channels} makes in particular the walker stay, with probability $\pi_1$, at his position during a \emph{finite}, i.e., \emph{non-vanishing}  amount of time $\Delta t = \epsilon$, that is, the values of $\varphi_L$ and $\varphi_R$ are irrelevant to this phenomenon.


\section{Numerical implementation of the Dirac Lindblad equation \eqref{eq:Lindblad_eq_2}}
\label{app:Num}

The Dirac Lindblad equation \eqref{eq:Lindblad_eq_2}, which can be written as Eq.\ \eqref{eq:struc}, is given, in position space, by
the system of equations \eqref{eq:sys1} and \eqref{eq:sys2}, which can be recast as
\begin{equation}
\partial_t \vec{r} + \mathbf{A} \partial_x \vec{r}  + \mathbf{A}' \partial_{x'} \vec{r}  = \mathbf{F} \vec{r}   \, ,
\end{equation}
where (i) $\vec{r} \equiv(r^0,r^1,r^2,r^3)^{\top}$ (see Eq.\ \eqref{eq:def}), (ii) the so-called Jacobian matrices are
\begin{subequations}
\begin{align}
\mathbf{A} &\equiv \mathrm{i}\mathbf{P} = \begin{bmatrix}
\cdot & \cdot & \cdot & -1   \\ 
\cdot & \cdot & \mathrm{i} & \cdot   \\ 
\cdot & -\mathrm{i} & \cdot & \cdot    \\ 
-1 & \cdot & \cdot & \cdot 
\end{bmatrix} ,  \\
\mathbf{A}' &\equiv -\mathrm{i}\mathbf{P}^{\dag} = \begin{bmatrix}
\cdot & \cdot & \cdot & -1   \\ 
\cdot & \cdot & -\mathrm{i} & \cdot  \\ 
\cdot & \mathrm{i} & \cdot & \cdot   \\ 
-1 & \cdot & \cdot & \cdot
\end{bmatrix} = \mathbf{A}^{\top} \, ,
\end{align}
\end{subequations}
where the dots stand for zeros and $\mathbf{P}$ is given in Eqs.\ \eqref{eq:PandQ}, and the so-called source-term matrix is, for $\gamma_1=0$,
\begin{equation}
\mathbf{F} \equiv \mathbf{Q}_{(\gamma_1=0,\gamma),m} = \begin{bmatrix}
\cdot & \cdot & \cdot  & \cdot \\ 
\cdot & \cdot & \cdot  & \cdot    \\ 
\cdot & \cdot     & -\gamma    & -2\,m   \\ 
\cdot & \cdot        &  2\,m      & -\gamma
\end{bmatrix} \, ,
\end{equation}
where $\mathbf{Q}$ is given in Eqs.\ \eqref{eq:PandQ}.
Given that the matrices $\mathbf{A}$ and $\mathbf{A}'$ are Hermitian and commute, there exists a unitary matrix,
\begin{equation}
\mathbf{U} \equiv \frac{1}{\sqrt{2}}\begin{bmatrix}
1      &    \cdot & \cdot           & 1   \\ 
\cdot          & \mathrm{i}       & 1          & \cdot   \\ 
\cdot          & -\mathrm{i}       & 1    & \cdot   \\ 
-1          & \cdot       &  \cdot      & 1
\end{bmatrix} \,  ,
\end{equation}
that simultaneously diagonalizes them,
\begin{subequations}
\begin{align}
\mathbf{\Lambda}&\equiv \mathbf{U} \mathbf{A} \mathbf{U}^{-1}=\text{diag}(-1,-1,1,1) \\
\mathbf{\Lambda}'&\equiv \mathbf{U} \mathbf{A}' \mathbf{U}^{-1}=\text{diag}(-1,1,-1,1) \, .
\end{align} 
\end{subequations}
This allows to rewrite the system of equations as
\begin{equation} \label{eq:full}
\partial_t \vec{v} + \mathbf{\Lambda} \partial_x \vec{v} + \mathbf{\Lambda}' \partial_{x'} \vec{v} = \mathbf{S} \vec{v}  \, ,
\end{equation}
where
\begin{subequations}
\begin{align}
\vec{v}\equiv(v^0,v^1,v^2,v^3)^{\top} &\equiv \mathbf{U} \vec{r} \\
\mathbf{S} &\equiv \mathbf{U} \mathbf{F} \mathbf{U}^{-1} \, .
\end{align}
\end{subequations}
System \eqref{eq:full} is a hyperbolic system of PDEs, and its solution is thus in particular solely determined by the initial condition $\vec{v}_{t=0}$.

We integrated this system numerically via the Strang operator-splitting method, which consists in splitting the single-time-step evolution into two parts: one corresponding to the homogeneous evolution of the system (i.e. $\mathbf{S}=0$), and the other one corresponding to the evolution with null fluxes (i.e. $\mathbf{\Lambda}=\mathbf{\Lambda}'=0$).
This method is particularly adapted to the present case, since the homogeneous solution is exactly solvable,
\begin{equation} \label{eq:hom}
	v^{\mu}_{t,x,x'}=v^{\mu}_{0, \, x-\lambda^{\mu} t, \, x'-{\lambda^{\mu}}'t} \, ,
\end{equation}
where the $\lambda^{\mu}$s (resp. ${\lambda^{\mu}}'$s), $\mu=0,...,3$, are the $4$ eigenvalues of the matrix $\mathbf{\Lambda}$ (resp. $\mathbf{\Lambda}'$).
For the second part of the evolution, one has to solve, as mentioned,
\begin{equation}
	\partial_t \vec{v} = \mathbf{S}\vec{v} \, .
\end{equation}
This equation has a well-known explicit solution, which requires the exponentiation of the matrix $\mathbf{S}$.
A direct numerical implementation of this exponential introduces well-known stiffness problems.
To address this issue, we implement instead the following first-order (i.e., Euler) explicit-implicit scheme,
\begin{equation}
\frac{\vec{v}_{t+\epsilon}-\vec{v}_t}{\epsilon} = \alpha \mathbf{S} \vec{v}_t + (1-\alpha) \mathbf{S} \vec{v}_{t+\epsilon} \, ,
\end{equation}
where the parameter $\alpha$ has been adjusted by hand to $\alpha=0.5$ \footnote{Our criterion to choose $\alpha$ has been to minimize, by hand, the discrepancy of the numerically obtained $\vec{v}_t$ from the discretized equation of motion, Eq.\ (\ref{eq:full}), at the final time of the simulation.}.

The accuracy of the splitting method can be improved from $O(\Delta t)$ to $O(\Delta t^2)$ by using the so-called Strang splitting, where we take half a step with the one-time-step source term evolution operator, $\mathbf{L}^{\mathrm{s}}_{\epsilon/2}$, a full step with the one-time-step homogeneous evolution operator $\mathbf{L}^{\text{h}}_{\epsilon}$, and finally half another step with the source term operator.
During a time interval $\epsilon$, the algorithm thus reads
\begin{equation}
	\vec{v}_{t+\epsilon} = \mathbf{L}^{\mathrm{s}}_{\epsilon/2}\mathbf{L}^{\text{h}}_{\epsilon} \mathbf{L}^{\mathrm{s}}_{\epsilon/2} \vec{v}_t \, .
\end{equation}

\section{About local noises} \label{app:local-noise}

A typical example of local noise, as defined by Eq.\ (\ref{eq:local-noise}), would be
\begin{equation} \label{eq:loc}
 \hat{V}_{(\lambda_{y})_{y}} = \hat{A} \hat{O}_{(\lambda_{y})_{y}} \hat{B} \, ,
\end{equation}
where the only operator, $\hat{O}_{(\lambda_{y})_{y}}$, that depends on the noise, (i) is purely local, i.e.,
\begin{equation}
\hat{O}_{(\lambda_{y})_{y}} \equiv \sum_z O^{z}_{(\lambda_{y})_{y}} \ket{z} \! \! \bra{z} \, ,
\end{equation}
with $O^{z}_{(\lambda_{y})_{y}}$ acting solely on the internal d.o.f. (which is why it has no hat), and (ii), depends locally on the noise, i.e., $O^{z}$ actually only depends on $\lambda_z$,
\begin{equation}
O^{z}_{(\lambda_{y})_{y}} \equiv o^{z}_{\lambda_z} \, .
\end{equation}
Indeed, if the random unitary has the form of Eq.\ (\ref{eq:loc}) with the precisions (i) and (ii) given just above, the matrix elements would be of the form of Eq.\ (\ref{eq:local-noise}), with
\begin{equation}
\hat{V}_{\lambda_z}^{xx'} \equiv \bra{x} \hat{A} \ket{z} o^{z}_{\lambda_z} \bra{z} \hat{B} \ket{x'} \, .
\end{equation}

Let us show that the fact that the operator $\hat{O}_{(\lambda_z)_z}$ is purely local is necessary for the noise to eventually be local without breaking the translational invariance of the system.
We do a reduction to absurd.
Assume that $\hat{O}_{(\lambda_z)_z}$ is not purely local, i.e., that, whatever\footnote{``Whatever'' is necessary, not only ``there is'', because of the requirement of translational invariance.} $x$, there exist $x'\neq x$ such that $O^{xx'}_{(\lambda_y)_y} \neq 0$.
Imagine now that $O^{xx'}_{(\lambda_y)_y}$ only depends on some $\lambda_z$, i.e.,  $O^{xx'}_{(\lambda_y)_y} = o^{xx'}_{\lambda_z}$.
 Translational invariance imposes that, whatever $x$, (i) $x-z = \text{cte}$ (we also have $x'-z = \text{cte}_2$, but we will not need it).
But, $x$ and $x'$ should not play a different role in a translationally invariant system, i.e., we must have $o^{xx'}_{\lambda_z} = o^{x'x}_{\lambda_z}$, so that (ii) $x'-z= \text{cte}$, which, together with (i), implies that $x'=x$, which contradicts the fact that $x'\neq x$, and completes the proof.
That  $\hat{O}_{(\lambda_z)_z}$ is purely local is thus necessary for the noise to be local. However, this condition is a priori not sufficient, one could imagine a unitary $O_{(\lambda_{y})_{y}}$ that acts purely locally, i.e., only on the internal degree of freedom, but whose action depends on the random variables at all points, i.e., depends indeed on the whole family $(\lambda_{y})_{y}$.

Notice that $\hat{V}^{\text{QW}}_{(\lambda_{y})_{y}}$, Eq.\ (\ref{eq:random-unit}), has a decomposition of the form of Eq.\ (\ref{eq:loc}) in which, more particularly, $\hat{A}$ is the identity, so that the sum over $z$ in Eq.\ (\ref{eq:local-noise}) reduces to the single term $z=x$, namely,
\begin{equation}
\bra{x}  \hat{V}^{\text{QW}}_{\lambda_x} \ket{x'}  \equiv  C'_{{\lambda_x}} \bra{x} \hat{S}  \ket{x'}  \, ,
\end{equation}
where
\begin{equation}
C'_{{\lambda_x}} \equiv C_{\epsilon {\bar{f}}_{t,x} +{\lambda_x}} \, .
\end{equation}

\section{Continuum limit of discrete-time quantum walks with temporal coin noise depending smoothly on the position}
\label{app:spatiotemp}

If, for each evolution $t \rightarrow t +\epsilon$, \emph{all} sequences $(\lambda_x^l)_x$ involved in the integral of Eq.\ \eqref{eq:QW-spatiotemp}, correspond, \emph{not} to outcomes of spatially-dependent random variables, but to values taken  by differentiable functions of $x$ (and with which they coincide in the continuum limit), then the continuum limit of each of them exists, and is obtained via Taylor expansion in $\epsilon$ of the spacetime-lattice dynamics.

We know, from Eq.\ (\ref{eq:Taylor1}), that the following Taylor expansion holds,
\begin{align} \label{eq:Cx}
&C'_{ \sqrt{\epsilon}\tilde{\lambda}_x} \bra{x} \hat{S} =  \\
&\ \ \ \,  1-\mathrm{i}\epsilon \bra{x}  \hat{H}^{o}_{t} \nonumber  \\
&\ \ \ \ \ + \bigg[  \mathrm{i} \sqrt{\epsilon} \sum_{l=0}^2  \tilde{\lambda}^l_x {L}_{l} \nonumber   \\
&\ \ \ \ \    - \frac{1}{2} (\sqrt{\epsilon})^2 \sum_{l=0}^2  (\tilde{\lambda}^l_x)^2 (L_{l})^2 \nonumber \\
&\ \ \ \ \   - (\sqrt{\epsilon})^2 \left[  \tilde{\lambda}^0_x ( \tilde{\lambda}^1_x \sigma^3 + \tilde{\lambda}^2_x \sigma^1) +  \tilde{\lambda}^2_x \tilde{\lambda}^3_x (\mathrm{i} \sigma^2) \right] \nonumber \\
&\ \ \ \ \   + O(\epsilon^{3/2}) \bigg] \bra{x} \nonumber \, .
\end{align} 
Similarly, we have
\begin{align} \label{eq:Cxprime}
&\hat{S}  \ket{{x'}} C'_{\sqrt{\epsilon}\tilde{\lambda}_{x'}} =  \\
& \ \ \ \ \ \ \ 1 + \mathrm{i}\epsilon \hat{H}^{o}_{t}  \ket{{x'}}  \nonumber  \\
& \! \! \!  + \ket{{x'}} \bigg[ - \mathrm{i} \sqrt{\epsilon} \sum_{l=0}^2  \tilde{\lambda}^l_{x'} {L}_{l} \nonumber   \\
&\ \ \ \ \ \ \ \ \,  - \frac{1}{2} (\sqrt{\epsilon})^2 \sum_{l=0}^2  (\tilde{\lambda}^l_{x'})^2 (L_{l})^2 \nonumber \\
&\ \ \ \ \  \  \ \ \, - (\sqrt{\epsilon})^2 \left[  \tilde{\lambda}^0_{x'} ( \tilde{\lambda}^1_{x'} \sigma^3 + \tilde{\lambda}^2_{x'} \sigma^1) +  \tilde{\lambda}^2_{x'} \tilde{\lambda}^3_{x'} (\mathrm{i} \sigma^2) \right] \nonumber \\
&\ \ \ \ \ \ \ \ \,  + O(\epsilon^{3/2}) \bigg]  \nonumber \, ,
\end{align} 

Plugging Eqs.\ (\ref{eq:Cx}) and (\ref{eq:Cxprime}) into Eq.\ (\ref{eq:Tay}) yields
{\small
\begin{align}
&\rho^{xx'} + \epsilon \partial_t \rho^{xx'} = \\
& \rho^{xx'} + \epsilon  \Big[ -\mathrm{i} \bra{x} [\hat{H}^o, \hat{\rho} ] \ket{x'} +  \mathcal{F}_{xx'}(\rho^{xx'}) \big] \nonumber  \\
& \ \ \ \   \ \,  + O(\epsilon^{3/2}) \, , \nonumber 
\end{align} }
where we recognize a standard Hamiltonian part, and where the noise term is given by
\begin{equation}
\mathcal{F}_{xx'}(\rho^{xx'}) \equiv \sum_{l=0}^2 \mathcal{F}^l_{xx'}(\rho^{xx'}) \, ,
\end{equation} 
where -- using that $\tilde{p}_{x,x'}^{l,(2)}(\tilde{\lambda}^l_x, \tilde{\lambda}^l_{x'})$ is symmetric (for the functions $v^l:x\mapsto v^l_x$ and $v^l:x'\mapsto v^l_{x'}$, defined below, to indeed be the same), which is the feature (\ref{eq:trans_inv3}) of the translationally invariant noise defined in Eqs.\ (\ref{eq:trans_inv}) -- we obtain
{\small
\begin{equation}
\mathcal{F}^l_{xx'}(\rho^{xx'}) \equiv c^l_{x,x'} L_l \rho^{xx'} L_l^{\dag} - \frac{1}{2} \left( v^l_x L_l^{\dag} L_l \rho^{xx'} + v^l_{x'} \rho^{xx'}  L_l^{\dag} L_l \right) \, ,
\end{equation}}
where the $L_l$s are given  by Eqs.\ (\ref{eq:Lls}) with $L_0 \equiv \mathbf{1}_2$, and with the 2-point ``correlation functions'' and the ``variances'' respectively given by
\begin{subequations}
\begin{align}
c^l_{x,x'} &\equiv \int \! \! d\tilde{\lambda}_x d\tilde{\lambda}_{x'} \ \tilde{p}_{x,x'}^{l,(2)}(\tilde{\lambda}^l_x, \tilde{\lambda}^l_{x'})  \ \tilde{\lambda}^l_x \tilde{\lambda}^l_{x'} \\
v^l_y &\equiv \int \! \! d\tilde{\lambda}_y \  \tilde{p}_{y}^{l,(1)}(\tilde{\lambda}^l_y) \ (\lambda_y^l)^2 \, .
\end{align}
\end{subequations}
Now using (i) that $\tilde{p}_{y}^{l,(1)}$ does not depend on $y$ -- which is the feature (\ref{eq:trans_inv1}) of the translationally invariant noise defined in Eqs.\ (\ref{eq:trans_inv}) --, so that $v_y = v$, and (ii) assuming that $c^l_{x,x'}$ actually only depends on $|x-x'|$ and not $(x,x')$, which is guaranteed if $\tilde{p}_{x,x'}^{l,(2)}$ behaves the same (feature (\ref{eq:trans_inv2}) of the translationally invariant noise defined in Eqs.\ (\ref{eq:trans_inv})), we obtain
{\small
\begin{align} \label{eq:final_spatial2}
\partial_t \rho^{xx'} =  -\mathrm{i} \bra{x} [\hat{H}^o, \hat{\rho} ] \ket{x'} +  \mathcal{N}_{\Gamma/2, \kappa_{|x-x'|}}(\rho^{xx'}) \, ,
\end{align} }
where the noise term is now
\begin{equation}
\mathcal{N}_{\Gamma/2, \kappa_{|x-x'|}}(\rho^{xx'}) \equiv \sum_{l=0}^2 \gamma_l \mathcal{M}^l_{\kappa^l_{|x-x'|}}(\rho^{xx'}) \, ,
\end{equation}
where we have renamed the ``variances'' in accordance with Sec.\ \ref{subsec:random-unitaries},
\begin{equation}
\gamma_l \equiv 2 v^l \, ,
\end{equation}
and introduced the ``correlation coefficients'',
\begin{equation}
\kappa^l_{|x-x'|} \equiv \frac{c^l_{x,x'}}{v^l} \in [-1,1] \, ,
\end{equation}
so that
\begin{equation}
\mathcal{M}^l_{\kappa^l_{|x-x'|}}(\rho^{xx'})  \equiv  \kappa^l_{|x-x'|} L_l \rho^{xx'} L_l - \frac{1}{2}\{ L_l^{\dag} L_l ,\rho^{xx'} \} \, .
\end{equation}

\end{document}